\documentclass{article}
\usepackage{arxiv}

\usepackage[utf8]{inputenc}
\usepackage{amsmath}
\usepackage{url}
\usepackage{hyperref}
\usepackage[numbers,sort]{natbib}  
\usepackage{amsfonts}
\usepackage{graphicx}
\usepackage{algorithm}
\usepackage{algpseudocode}
\usepackage{subfig}
\usepackage{float}
\usepackage{booktabs}
\usepackage{mathtools}
\usepackage{etoolbox}

\usepackage{caption}

\newcount\Comments  
\usepackage{color}
\definecolor{darkgreen}{rgb}{0,0.5,0}
\definecolor{purple}{rgb}{1,0,1}
\newcommand{\kibitz}[2]{\ifnum\Comments=0\textcolor{#1}{#2}\fi}

\title{Robust Tensor Recovery with Fiber Outliers for Traffic Events}

\author{
  Yue Hu\\
  Vanderbilt University\\
  Nashville, TN37212 \\
  \texttt{yue.hu@vanderbilt.edu} \\
   \And
 Daniel Work \\
  Vanderbilt University\\
  Nashville, TN37212 \\
  \texttt{dan.work@vanderbilt.edu} 
 }
 
\begin{document}
\maketitle

\begin{abstract}
Event detection is gaining increasing attention in smart cities research. Large-scale mobility data serves as an important tool to uncover the dynamics of urban transportation systems, and more often than not the dataset is incomplete. In this article, we develop a method to detect extreme events in large traffic datasets, and to impute missing data during regular conditions. Specifically, we propose a robust tensor recovery problem to recover low rank tensors under fiber-sparse corruptions with partial observations, and use it to identify events, and impute missing data under typical conditions. Our approach is scalable to large urban areas, taking full advantage of the spatio-temporal correlations in traffic patterns. We develop an efficient algorithm to solve the tensor recovery problem based on the \textit{alternating direction method of multipliers} (ADMM) framework. Compared with existing $l_1$ norm regularized tensor decomposition methods, our algorithm can exactly recover the values of uncorrupted fibers of a low rank tensor and find the positions of corrupted fibers under mild conditions. Numerical experiments illustrate that our algorithm can exactly detect outliers even with missing data rates as high as 40\%, conditioned on the outlier corruption rate and the Tucker rank of the low rank tensor. Finally, we apply our method on a real traffic dataset corresponding to downtown Nashville, TN, USA and successfully detect the events like severe car crashes, construction lane closures, and other  large events that cause significant traffic disruptions.
\end{abstract}


\section{Introduction}
\subsection{Motivation}\label{subsect:motivation}
Event detection is an increasing interest in urban studies~\cite{liu2014anomaly,wen2018event}. Efficiently analyzing the impact of large events can help us assess the performance of urban infrastructure and aid urban management. Nowadays, with the development of intelligent transportation systems, large scale traffic data is accumulating via loop detectors, GPS, high-resolution cameras, etc. The large amount of data provides us insight into the dynamics of urban environment in face of large scale events. The main objective of this article is to develop a method to detect extreme events in traffic
datasets describing large urban areas, and to impute missing data during regular conditions. 

There are two major challenges in event detecting outliers in traffic datasets. Firstly, most large traffic datasets are incomplete~\cite{conklin2002use,chen2012retrieval,tan2013tensor}, meaning there are a large number of entries for which the current traffic condition is not known. The missing data can be caused by the lack of measurements (e.g., no instrumented vehicles recently drove over the road segment), or due to senor failure (e.g., a traffic sensor which loses communication, power, is physically damaged). Missing data can heavily influence the performance of traffic estimation~\cite{chen2001study,chen2012retrieval,williams2003modeling,tan2013tensor}, especially as the missing data rate increases. Naive imputation of the missing entries to create a complete dataset is problematic, because without a clear understanding of the overall pattern, an incorrect value can be imputed that will later degrade the performance of an outlier detection algorithm. Consequently, missing data should be carefully handled.

The second challenge is to fully capture and utilize the pattern of regular traffic, in order to correctly impute missing data and separate the outliers out from regular traffic. Studies have suggested that for regular traffic patterns, there exist systematic correlations in time and space~\cite{yang2014detecting, asif2016matrix,furtlehner2010spatial,han2011analysis}. For example, due to daily commute patterns, traffic conditions during Monday morning rush hour are generally repeated but with small variation from one week to the next (e.g., the rush hour might start a little earlier or last a little longer). Also, traffic conditions are spatially structured due to the network connectivity, ending up in global patterns. For example, the traffic volume on one road segment should influence and be influenced by its down stream and upper stream traffic volume. 

Most existing researches have not fully addressed these two challenges. On the one hand, missing data and outlier detection tend to be dealt with separately. Either it is assumed that the dataset is complete, for the purpose of outlier detection~\cite{hodge2004survey,gupta2013outlier}, or it is assumed that the dataset is clean, for the purpose of missing data imputation~\cite{qu2008bpca,afifi1966missing}. Yet in reality missing data and outliers often exist at the same time. On the other hand, currently most studies on traffic outliers consider only a single monitor spot or a small region~\cite{ guo2015real,chen2010comparison,park2003empirical,turochy2000applying}, not fully exploiting the spatio-temporal correlations. Only a few studies scale up to large regions to capture the urban-wise correlation, including the work of Yang et al.~\cite{yang2014detecting} proposing a coupled Bayesian \textit{robust principal component analysis} (robust PCA or RPCA) approach to detect road traffic events, and Liu et al.~\cite{liu2011discovering} constructing a spatio-temporal outlier tree to discover the causal interactions among outliers. 

In this article, we tackle the traffic outlier event detection problem from a different perspective, taking into account missing data and spatio-temporal correlation. Specifically, we model a robust tensor decomposition problem, as illustrated in subsection~\ref{subsection:Sol}. Furthermore, we develop an efficient algorithm to solve it. We note that the application of our method is not limited to traffic event detection, but can also be applied to general pattern recognition and anomaly detection, where there exist multi-way correlations in the dataset, and in either full or partial observations.

\subsection{Solution approach}\label{subsection:Sol}
In this subsection, we develop the robust tensor decomposition model for traffic extreme event detection problem, and develop a robust tensor completion problem to take partial observation  into account.

\begin{figure}
\centering
\includegraphics[width=0.4\linewidth]{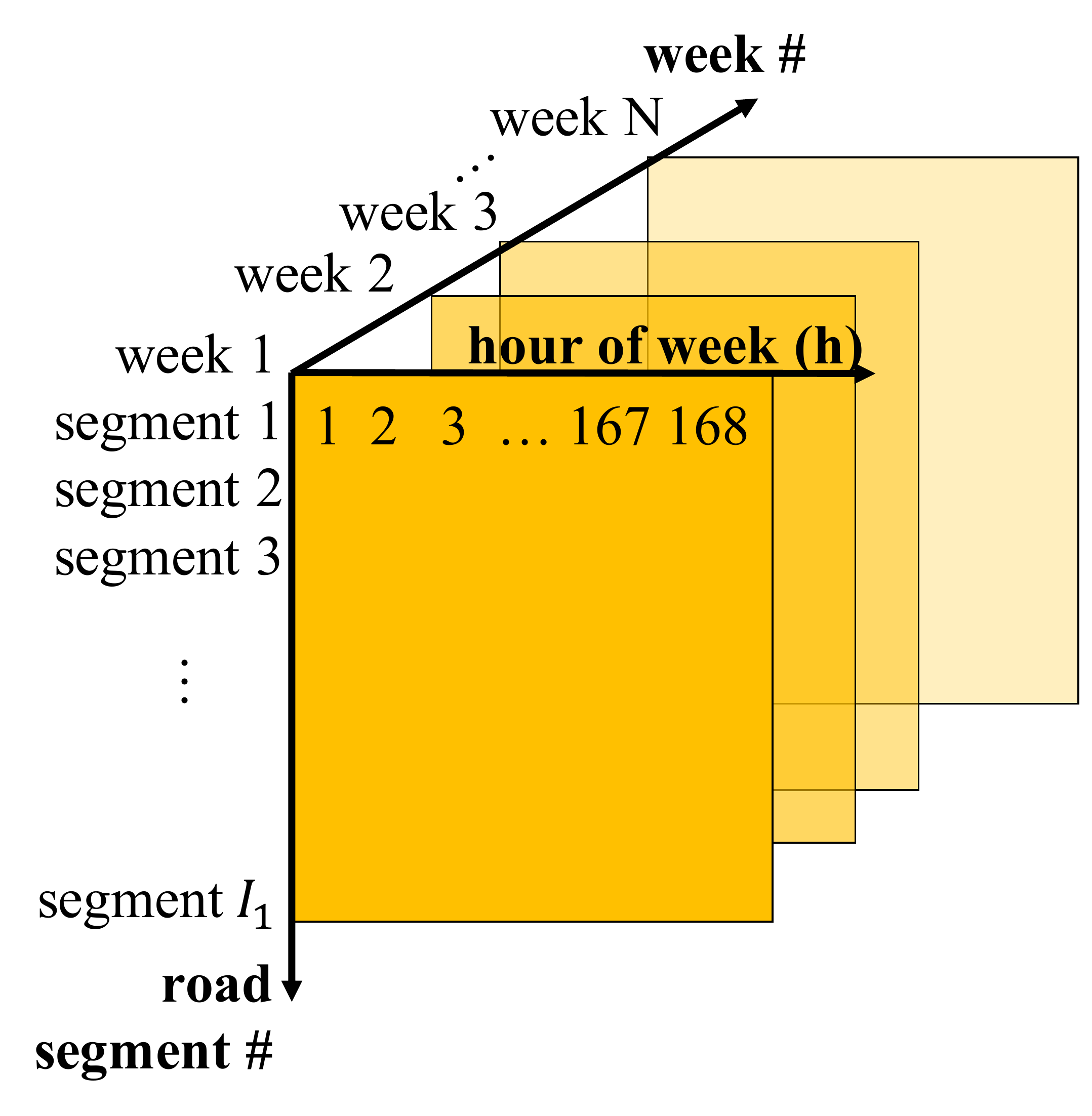}
\caption{ Traffic data is arranged in three-way tensor, with the dimensions corresponding to \textit{i}) the hour of the week, \textit{ii}) the road segment number, and \textit{iii}) the week number.}
\label{fig:3way}
\end{figure}

To exploit these temporal and spatial structures, a tensor~\cite{kolda2009tensor,goldfarb2014robust} is introduced to represent the traffic data over time and space. We form a three-way tensor, as shown in Figure \ref{fig:3way}. The first dimension is the road segment, the second dimension is the time of the week, (Monday Midnight-1am all the way to Sunday 11pm-midnight, $24 \times 7 = 168$ entries in total), and the third dimension is the week in the dataset. In this way, the temporal and spacial patterns along different dimensions are naturally encoded. One effective way to quantify this multi-dimensional correlation is the Tucker rank of the tensor~\cite{kolda2009tensor,goldfarb2014robust,tucker1966some}, which is the generalization of matrix rank to higher dimensions. 

As for the extreme events, we expect them to occur relatively infrequently. We encode the outliers in a sparse tensor, which is organized in the same way as the traffic data tensor. Furthermore, extreme events tend to affect the overall traffic of an urban area. That is, we assume that at the time when extreme event occurs, the traffic data of all road segments deviates from the normal pattern. Thus, the outliers occur sparsely as fibers along the road segment dimension with hour of week fixed. This sets it apart from random noises, which appear scattered over the whole tensor entries and unstructured. This fiber-wise sparsity problem is studied in 2D matrix cases~\cite{xu2010robust}, where the $l_{2,1}$ norm is used to control the column sparsity, and we adapt it for higher dimensions.

\begin{figure}
\centering
\includegraphics[width=\linewidth]{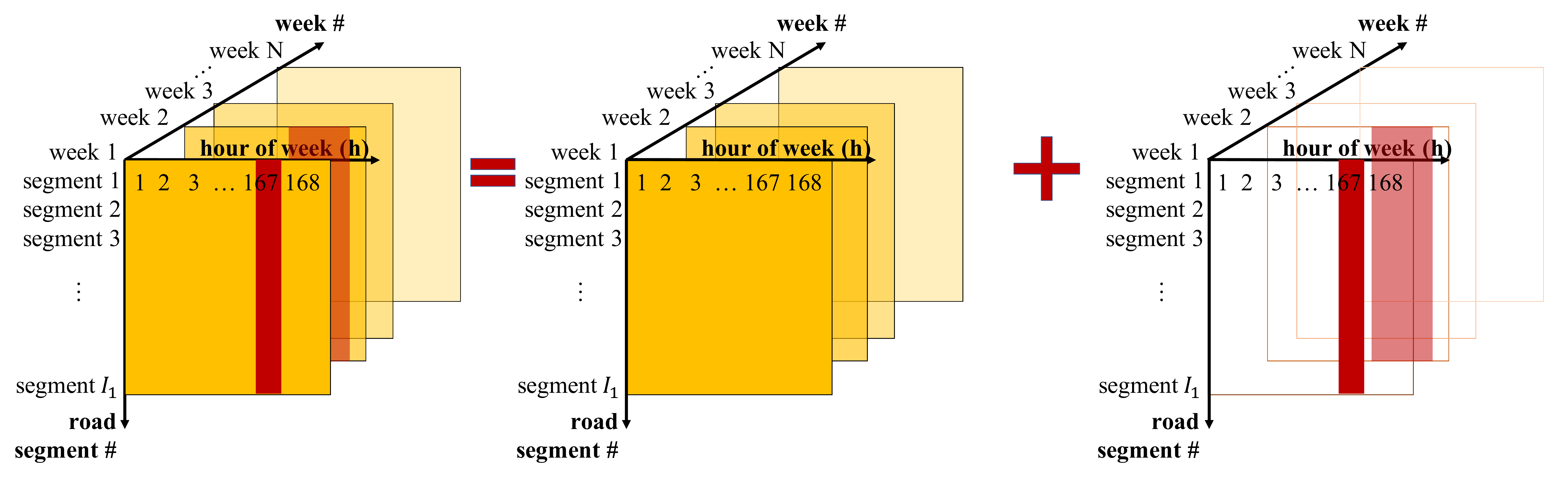}
\caption{Observation data decomposed into low rank tensor for regular traffic and fiber-sparse tensor for outlier events.}
\label{fig:decom_illu}
\end{figure}

Putting these together, we organize the traffic data into a tensor $\mathcal{B}$, then decompose it into two parts, 
\begin{equation*}
     \mathcal{B = X + E,}
\end{equation*}
where  tensor $\mathcal{X}$ contains the data describing the regular traffic patterns, and tensor $\mathcal{E}$ denotes the outliers, as illustrated in Figure~\ref{fig:decom_illu}. Because the normal traffic patterns is assumed to have strong correlation in time and space, it is approximated by a low rank tensor $\mathcal{X}$, Similarly, because outliers are infrequent, we expect the tensor containing outliers, $\mathcal{E}$, to be sparse. With these ideas in mind, it is possible formulate the following optimization problem:  
\begin{equation}
\label{eq:orig_form0}
\begin{aligned}
    &\underset{\mathcal{X,E}}{\text{min}}
    & &\text{rank}(\mathcal{X})+ \lambda  \text{  sparsity} (\mathcal{E})\\
    & \text{s.t.}
    & & \mathcal{B = X + E.}
    \end{aligned}
\end{equation}
The objective function in problem \eqref{eq:orig_form0} is the weighted cost of tensor rank of $\mathcal{X}$ denoted as rank($\mathcal{X}$), the fiber-wise sparsity of $\mathcal{E}$ is denoted as sparsity($\mathcal{E}$), and $\lambda$ is a weighting parameter balancing the two costs. A more precise formulation of the problem is provided in Section~\ref{Sec:Method}. 

In the presence of missing data, we require the decomposition to match the observation data only at the available entries, and come to the optimization problem:
\begin{equation} 
\label{eq:TC0}
\begin{aligned}
    & \underset{\mathcal{X,E}}{\text{min}}
    & &\text{rank}(\mathcal{X}) + \lambda  \text{  sparsity} (\mathcal{E})\\ 
    & \text{s.t.}
    & & \mathcal{B}_{i_1i_2\dots i_N}= \mathcal{(X+E)}_{i_1i_2\dots i_N}, \\
    & &&\text{where } (i_1,i_2,\dots,i_N) \text{ is an observed entry. }
\end{aligned}
\end{equation}

In this paper, we turn problem~\eqref{eq:orig_form0} and~\eqref{eq:TC0} into convex programming problems, and solve them by extending from matrix case to tensor case a singular value thresholding algorithm~\cite{cai2010singular,candes2011robust} based on  \textit{alternating direction method of multipliers} (ADMM) framework. Our algorithm can exactly recover the values of uncorrupted fibers of the low rank tensor, and find the positions of corrupted fibers, based on relatively mild condition of observation and corruption ratio.\footnote{The resulting source code is available at \url{https://github.com/Lab-Work/Robust_tensor_recovery_for_traffic_events}.}

\subsection{Contributions and outline}
To summarize, this work has three main contributions:
\begin{enumerate}
    \item  We propose a new robust tensor recovery with fiber outliers model for traffic extreme event detection under full or partial observations, to take full advantage of spatial-temporal correlations in traffic pattern.
    \item We propose ADMM based algorithms to solve the robust tensor recovery under fiber-sparse corruption. Our algorithm can exactly recover the values of uncorrupted fibers of the low rank tensor, and find the positions of corrupted fibers under mild conditions.
    \item We apply our method on a large traffic dataset in downtown Nashville and successfully detect large events.
\end{enumerate}

The rest of the paper is organized as follows. In Section~\ref{Sec:Prelim}, we provide a review of tensor basics and related robust PCA methods. In Section~\ref{Sec:Method}, we formulate the tensor outlier detection problem, and propose efficient algorithms to solve it.  In Section~\ref{Sec:Simulation}, we demonstrate the effectiveness of our method by numerical experiments. In Section~\ref{Sec:case}, we apply our method on real world data set and show its ability to find large scale events. 

\section{Related work}
In this section, we describe literature on outlier detection. We also compare our methodology with other relevant works.

\subsection{outlier detection}
The outliers we are interested in this work are due to outliers caused by extreme events. Another related problem considers methods to detect outliers caused by data measurement errors, such as sensor malfunction, malicious tampering, or measurement error~\cite{boto2010wavelet,chen2010comparison,park2003empirical}. The latter methods can be seen as a part of a standard data cleaning or data pre-processing step.  On the other hand, outliers caused by extreme traffic have valuable information for congestion management, and can provide agencies with insights into the performance of urban network. The works~\cite{kong2018lotad,pang2013detection,liu2011discovering} explore the problem of outlier detection caused by events, while the works~\cite{liu2014anomaly,sun2017dxnat,xu2013identifying,kwoczek2014predicting} focus on determining the root causes of the outlier.

\subsection{Low rank matrix and tensor learning}
Low rank matrix and tensor learning has been widely used to utilize the inner structure of the data. Various application have benefited from matrix and tensor based methods, including data completion~\cite{song2019tensor,asif2016matrix}, link prediction~\cite{dunlavy2011temporal}, network structure clustering~\cite{wang2014discovering}, etc. 

The most relevant works with ours are robust matrix and tensor PCA for outlier detection. $l_1$ norm regularized robust tensor recovery, as proposed by Goldfarb and Qin~\cite{goldfarb2014robust}, is useful when data is polluted with unstructured random noises. Tan et al.~\cite{tan2013traffic} also used $l_1$ norm regularized tensor decomposition for traffic data recovery, in face of random noise corruption. But if outliers are structured, for example grouped in columns, $l_1$ norm regularization does not yield good results. In addition, although traffic is also modeled in tensor format in~\cite{tan2013traffic}, only a single road segment is considered, not taking into account network spacial structures.

In face of large events, outliers tend to group in columns or fibers in the dataset, as illustrated in section~\ref{subsection:Sol}. $l_{2,1}$ norm regularized decomposition is suitable for group outlier detection, as shown in~\cite{tang2011robust,xu2010robust} for matrices, and~\cite{zhou2017outlier,ren2016robust} for tensors. In addition, Li et al.~\cite{li2018multi} introduced a multi-view low-rank analysis framework for outlier detection, and Wen et al.~\cite{wen2018event} used discriminant tensor factorization for event analytics. Our methods differ from the existing tensor outliers pursuit~\cite{zhou2017outlier,ren2016robust} in that they are dealing with slab outliers, i.e., outliers form an entire slice instead of fibers of the tensor. Moreover, compared with existing works, we take one step further and deal with partial observations. As stated in Section \ref{subsect:motivation}, without an overall understanding of the underlying pattern, we can easily impute the missing entries incorrectly and influence our decision about outliers. We will show in  Section \ref{subsec:TC_sim} simulation that our new algorithm can exactly detect the outliers even with 40\% missing rate, conditioned on the outlier corruption rate and the rank of the low rank tensor.

\section{Preliminaries} \label{Sec:Prelim}

In this section, we briefly review the mathematical preliminaries for tensor factorization, adopting the notation of Kolda and Bader~\cite{kolda2009tensor}, and Goldfarb and Qin~\cite{goldfarb2014robust}. We also summarize robust PCA~\cite{candes2011robust}, since it serves as a foundation for our extension to higher-order tensor decomposition. 

 \subsection{Tensor basics}
 In this article, a tensor is denoted by an Euler script letter (e.g., $\mathcal{X}$); a matrix by a boldface capital letter (e.g., $\mathbf{X}$); a vector by a boldface lowercase letter (e.g., $\mathbf{x}$); and a scalar by a lowercase letter (e.g., $x$). A tensor of order $N$ has $N$ dimensions, and can be equivalently described as an $N$-way tensor or an $N$-mode tensor. Thus, matrix is a second order tensor, and vector is a first order tensor.

A \textit{fiber} is a column vector formed by fixing all indices of a tensor but one. In a matrix for example, each column can be viewed as a mode-one fiber, and each row a mode-two fiber. 

The \textit{unfolding} of a tensor $\mathcal{X}$ in the $n^{\text{th}}$ mode results in a matrix $\mathbf{X}_{(n)}$, which is formed by rearranging the mode-$n$ fibers as its columns. This process is also called flattening or matricization. The inverse function of unfolding is denoted as $\text{fold}_n(\cdot )$, i.e.,
\begin{equation*}\label{eq:fold}
    \text{fold}_n(\mathbf{X}_{(n)} )= \mathcal{X}.
\end{equation*}

The \textit{inner product} of two tensors $\mathcal{X,Y} \in \mathbb{R}^{I_1 \times I_2 \times \dots \times I_N}$ is the sum of their element-wise product, similar to vector inner products. Let $x_{i_1i_2\dots i_N}$ and $y_{i_1i_2\dots i_N}$ denote the $(i_1,i_2, \cdots, i_N)$ element of $\mathcal{X}$ and $\mathcal{Y}$ respectively. Then tensor inner product can be expressed as
\begin{equation*}
    \langle \mathcal{X,Y} \rangle = \sum_{i_1=1}^{I_1}\sum_{i_2=1}^{I_2} \dots \sum_{i_N=1}^{I_N}x_{i_1i_2\dots i_N}y_{i_1i_2\dots i_N}.
\end{equation*}

The \textit{tensor Frobenius norm} is denoted by $\| \cdot \|_F$, and computed as
$$\|\mathcal{X}\|_F = \sqrt{\langle\mathcal{X},\mathcal{X}\rangle}.$$

\textit{Multiplication} of a tensor by a matrix in mode $n$ is performed by multiplying every mode-$n$ fiber of the tensor by the matrix. The mode-$n$ product of a tensor $\mathcal{X}\in \mathbb{R}^{I_1 \times I_2 \times\dots \times I_N}$ and a matrix $\mathbf{A} \in \mathbb{R}^{J \times I_n}$ is denoted by $\mathcal{X} \times_n \mathbf{A} = \mathcal{Y}$, where $\mathcal{Y} \in \mathbb{R}^{I_1 \times I_2 \times \dots  \times I_{n-1} \times J \times I_{n+1} \times\dots \times I_N}$. It is also equivalently written via its mode-$n$ unfolding as $\mathbf{Y}_{(n)} := \mathbf{AX}_{(n)}$.

The \textit{Tucker decomposition}~\cite{goldfarb2014robust,kolda2009tensor} is the generalization of matrix PCA in higher dimensions. It approximates a tensor  $\mathcal{X} \in \mathbb{R}^{I_1 \times I_2 \times\dots \times I_N}$ as a core tensor multiplied in each mode $n$ by an appropriately sized matrix $\mathbf{U}^{(n)}$:
\begin{equation}\label{eq:TuckerDC}
    \mathcal{X} \approx \mathcal{G} \times_1 \mathbf{U}^{(1)}\times_2 \mathbf{U}^{(2)}\times\dots \times_N \mathbf{U}^{(N)}.
\end{equation}
$\mathcal{G} \in \mathbb{R}^{c_1 \times c_2 \times\dots \times c_N}$ in~\eqref{eq:TuckerDC} is called the core tensor, where $c_1$ through $c_N$ are given integers. If $c_n < I_n$ for some $n$ in $(1,2, \dots, N)$,  the core tensor $\mathcal{G}$ can be viewed as a compressed version of $\mathcal{X}$. The matrices $\mathbf{U}^{(n)} \in \mathbb{R} ^{I_n \times c_n}$ are factor matrices, which are usually assumed to be orthogonal. 

The \textit{n-rank} of $\mathcal{X}$, denoted by $\text{rank}_n\mathcal{(X)}$, is the column rank of $\mathbf{X}_{(n)}$. In other words, it is the dimension of the vector space spanned by the mode-$n$ fibers. If we denote the $n$-rank of the tensor $\mathcal{X}$ as $R_n$ for $n = 1, 2, \dots, N$, i.e., $R_n = \text{rank}_n\mathcal{(X)}$, then the set of the $N$ $n$-ranks of  $\mathcal{X}$, $(R_1, R_2, \dots, R_N)$, is called the \textit{Tucker Rank}~\cite{kolda2009tensor}.  In Tucker decomposition~\eqref{eq:TuckerDC}, if $c_n =\text{rank}_n\mathcal{(X)} $ for all $n$ in $(1,2, \dots, N)$, then the Tucker decomposition is exact. In this case, we can easily conduct the decomposition by setting  $\mathbf{U}^{(n)}$ as the  left singular matrix of $\mathbf{X}_{(n)}$. Otherwise, if $c_n < \text{rank}_n\mathcal{(X)} $, the decomposition holds only as an approximation, and will be harder to be solved.

\subsection{Robust PCA}
We briefly summarize robust variants of PCA in the matrix setting, which are extended to higher-order tensor settings in Section~\ref{Sec:Method}.
Robust PCA belongs to the family of dimension-reduction methods aiming at combating the so-called \textit{curse of dimensionality} that often appears when dealing with large, high dimensional datasets, by finding the best representing low-dimensional subspace. PCA is a widely used technique in this family, yet it is sensitive to corruptions~\cite{candes2011robust}.  For example, if we have a large data matrix that comes from a low rank matrix randomly corrupted by large noises, i.e., 
$$ \mathbf{B = X + E},$$
where $\mathbf{B} \in \mathbb{R}^{I_1 \times I_2}$ is the data matrix, $\mathbf{X} \in \mathbb{R}^{I_1 \times I_2}$ is a low rank matrix, and $\mathbf{E} \in \mathbb{R}^{I_1 \times I_2}$ is a sparse corruption matrix of arbitrary magnitude. In this setting traditional PCA fails at finding the subspace for $\mathbf{X}$ given only $\mathbf{B}$. 

To address the problem of gross corruption, Cand{\`e}s et al.~\cite{candes2011robust} proposed a RPCA method known as  \textit{Principle Component Pursuit} (PCP):

\begin{equation}
\label{eq:CandesPCP}
\begin{aligned}
    & \underset{\mathbf{X,E}}{\text{min}}
    & & \|\mathbf{X}\|_*+ \lambda \|\mathbf{E}\|_1\\
     & \text{s.t.}
    & & \mathbf{B = X+E,}
\end{aligned}
\end{equation}
with the $l_1$ matrix norm $\| \cdot\|_1$ of $\mathbf{E}$ given by: 
\begin{equation*}
    \|\mathbf{E}\|_1 := \sum_{i= 1}^{I_1}\sum_{j=1}^{I_2}|e_{ij}|,
\end{equation*}
and where $e_{i,j}$ denotes the $(i,j)$th element of $\mathbf{E}$. The nuclear norm of a matrix $\mathbf{X}$ is denoted as $\| \cdot \|_*$ and is computed as the sum of the singular values of $\mathbf{X}$:
\begin{equation*}
    \|\mathbf{X}\|_* := \sum_i\sigma_i.
\end{equation*}
 where the SVD of $\mathbf{X}$ is $\mathbf{X} = \sum_i\sigma_i\mathbf{u}_i\mathbf{v}_i^T$. 
 
 The nuclear norm in~\eqref{eq:CandesPCP} is proposed as the tightest convex relaxation of matrix rank~\cite{cai2010singular}; and  the $l_1$ norm is the convex approximation for element-wise matrix sparsity. Cand{\`e}s et al.~\cite{candes2011robust} showed that PCP can exactly recover a low rank matrix when it is grossly corrupted at sparse entries. Moreover, by adopting ADMM algorithm, it is possible to solve \eqref{eq:CandesPCP} in polynomial time. The PCP formulation~\eqref{eq:CandesPCP} requires incoherence of the column space of the sparse matrix~$\mathbf{E}$~\cite{candes2011robust,xu2010robust}, and does not address the setting where entire columns are corrupted. 
 
An alternative problem formulation using an $l_{2,1}$ norm on $\mathbf{E}$ in~\eqref{eq:CandesPCP} is introduced for matrix recovery with column-wise corruption~\cite{xu2010robust,tang2011robust}.  The $l_{2,1}$ norm of a matrix $\mathbf{E} \in \mathbb{R}^{I_1 \times I_2}$ is defined as
\begin{equation*}
    \|\mathbf{E}\|_{2,1} = \sum_{j=1}^{I_2}\sqrt{\sum_{i=1}^{I_1}(e_{ij})^2}.
\end{equation*}
It is essentially a form of group lasso~\cite{hastie2015statistical}, where each column is treated as a group. Minimizing $ \|\mathbf{E}\|_{2,1}$ encourages the entire columns of $\mathbf{E}$ to be zero or non-zero, and leads to fewer non-zero columns. 

Note that it is hard to recover an uncorrupted column from a completely corrupted one. Therefore, instead of trying to recover the complete low rank matrix, Xu et al.~\cite{xu2010robust} seeks instead to recover the exact low-dimensional subspace while identifying the location of the corrupted columns. Tang et al.~\cite{tang2011robust} makes an assumption that if a column is corrupted (i.e., $\mathbf{E}$ has nonzero entries in this column), then the entries of the corresponding column in the low-rank matrix $\mathbf{X}$ are zero. This choice allows exact recovery of the low rank matrix in the non-corrupted columns.

Like PCA for a matrix, we note that the Tucker decomposition of a tensor is also sensitive to gross corruption~\cite{goldfarb2014robust}. Motivated by the ideas of Cand{\`e}s et al.~\cite{candes2011robust} and Tang~\cite{tang2011robust} for robust matrix PCA, in the next section we address the problem of robust decomposition of tensors with gross fiber-wise corruption.

\section{Methods} \label{Sec:Method}
In this section, we define and pose the higher-order tensor decomposition problem in the presence of fiber outliers and its partial-observation variant as convex programs, and provide efficient algorithms to solve them. 

\subsection{Problem formulation} \label{Formulation}
The precise setup of the problem is as follows. We are given a high dimensional data tensor $ \mathcal{B} \in \mathbb{R}^{I_1 \times I_2 \times\dots \times I_N}$ which is composed of a low-rank tensor $\mathcal{X} \in \mathbb{R}^{I_1 \times I_2 \times\dots \times I_N}$ that is corrupted in a few fibers. In other words, we have $\mathcal{B = X +E}$, where $\mathcal{E}  \in \mathbb{R}^{I_1 \times I_2 \times\dots \times I_N}$ is the sparse fiber outlier tensor. We know neither the rank of $\mathcal{X}$, nor the number and position of non-zero entries of $\mathcal{E}$. Given only $\mathcal{B}$, our goal is to reconstruct $\mathcal{X}$ on the non-corrupted fibers, as well as identify the outlier location. Moreover, we might have only partial observations of $\mathcal{B}$, and we seek to complete the decomposition nevertheless.

We do assume knowledge of the mode along which the fiber outliers are distributed; without loss of generality let it be the first dimension. Then it is equivalent to say the mode-1 unfolding of the outlier tensor is column-wise sparse. Thus, we can formulate the problem as:
\begin{equation}
\label{eq:orig_form}
\begin{aligned}
    &\underset{\mathcal{X,E}}{\text{min}}
    & &\text{rank}(\mathcal{X})+ \lambda \|\mathbf{E}_{(1)}\|_{2,1}\\
    & \text{s.t.}
    & & \mathcal{B = X + E.}
    \end{aligned}
\end{equation}
Computing the rank of a tensor $\mathcal{X}$, denoted by rank($\mathcal{X}$) is generally an NP-hard problem~\cite{haastad1990tensor,goldfarb2014robust}. One commonly used convex relaxation of the tensor rank is $\sum_i\|\mathbf{X}_{(i)}\|_*$, which sums the nuclear norm of the tensor unfoldings in all modes~\cite{goldfarb2014robust}. In this way, we generalize the matrix nuclear norm to the higher-order case, and explore the potential low rank structure in all dimensions.
Problem \eqref{eq:orig_form} thus becomes
\begin{equation}\label{eq:HoPCA-C}
\begin{aligned}
    &\underset{\mathcal{X,E}}{\text{min}}
    & &\sum_{i=1}^N\|\mathbf{X}_{(i)}\|_*+\lambda \|\mathbf{E}_{(1)}\|_{2,1}\\
    & \text{s.t.}
    & & \mathcal{B = X + E.}
\end{aligned}
\end{equation}

Next, we deal with the case when the data is only partially available, in addition to observation data being grossly corrupted. We only know the entries $(i_1,i_2, \dots ,i_N) \in \Omega$, where $\Omega \subset [I_1] \times [I_2] \times \dots \times [I_N]$ is an observation index set. Let $\mathcal{X}_{\Omega}$ denote the projection of $\mathcal{X}$ onto the tensor subspace supported on $\Omega $. Then $\mathcal{X}_{\Omega}$ can be defined as
\begin{equation*}
  \mathcal{X}_{\Omega}=\begin{cases}
    \mathcal{X}_{i_1 i_2 \dots i_N}, & (i_1,i_2, \dots ,i_N) \in \Omega\\
    0, &  \text{otherwise.}
  \end{cases}
\end{equation*}
Then we can force the decomposition to match the observation data only at the available entries, and  find the decomposition that minimizes the weighted cost of tensor rank and sparsity, leading to the following model:
\begin{equation}\label{eq:TC} 
\begin{aligned}
    & \underset{\mathcal{X,E}}{\text{min}}
    & & \sum_{i=1}^N\|\mathbf{X}_{(i)}\|_*+ \lambda \|\mathbf{E}_{(1)}\|_{2,1}\\ 
    & \text{s.t.}
    & & \mathcal{B}_\Omega = \mathcal{(X+E)}_\Omega.
\end{aligned}
\end{equation}
Note that related problems to~\eqref{eq:TC} for the matrix setting are  addressed in~\cite{candes2011robust,chen2011robust}).

\subsection{Algorithm} \label{Sec:Algorithm}
In this section, we develop algorithm for tensor decomposition with fiber-wise corruption model formulated in Section~\ref{Formulation}. We first solve~\eqref{eq:HoPCA-C} for the full-observation setting, then for the partial-observation setting~\eqref{eq:TC}, adopting an ADMM method~\cite{goldfarb2014robust} for each. 

\subsubsection{Higher-order RPCA} \label{Higher-order RPCA}
Problem \eqref{eq:HoPCA-C} is difficult to solve because the terms $\|\mathbf{X}_{(i)}\|_*$ in the objective function are interdependent, since each $\mathbf{X}_{(i)}$ is unfolded from the same tensor $\mathcal{X}$. Alternatively, we split $\mathcal{X}$ into $N$ auxiliary variables, $\mathcal{X}_1, \mathcal{X}_2, \dots, \mathcal{X}_N \in \mathbb{R}^{I_1 \times I_2 \times\dots \times I_N}$, and rewrite \eqref{eq:HoPCA-C} as:

\begin{equation}
\label{eq:HoPCA-C-xn}
\begin{aligned}
& \underset{\mathcal{X}_i,\mathcal{E}}{\text{min}}
& & \sum_{i=1}^N\|\mathbf{X}_\mathnormal{i(i)}\|_*+ \lambda  \|\mathbf{E}_{(1)}\|_{2,1} \\
& \text{s.t.}
& & \mathcal{B = X_\mathnormal{i}+E}, i = 1,2,\dots,\ N,
\end{aligned}
\end{equation}
where $\mathbf{X}_{i(i)}$ are the unfoldings of $\mathcal{X}_i$ in the $i^\text{th}$ mode.
The $N$ constraints $\mathcal{B = X_\mathnormal{i}+E}$ ensure that $\mathcal{X}_1, \mathcal{X}_2, \dots, \mathcal{X}_N$ are all equal to the original $\mathcal{X}$ in problem~\eqref{eq:HoPCA-C}.

Next, we proceed to solve problem \eqref{eq:HoPCA-C-xn} via an ADMM algorithm. A full explanation of the general ADMM framework can be found in~\cite{boyd2011distributed}. The corresponding augmented Lagrangian function for problem \eqref{eq:HoPCA-C-xn} is 
\begin{equation*}
\begin{aligned}
    \mathcal{L}(\mathcal{X}_1, \mathcal{X}_2, \dots, \mathcal{X}_N,\mathcal{E}, \mathcal{Y}_1, \mathcal{Y}_2, \dots, \mathcal{Y}_N;\mu)=
    \sum_{i=1}^N\|\mathbf{X}_{i(i)}\|_*+ \lambda \|\mathbf{E}_{(1)}\|_{2,1}  + \\ \sum_{i=1}^N \left( \frac{\mu}{2} \|\mathcal{X_\mathnormal{i}+E-B)}\|_F^2-\langle \mathcal{Y_\mathnormal{i},X_\mathnormal{i}+E-B }\rangle\right).
\end{aligned}
\end{equation*}
Here $\mathcal{Y}_\mathnormal{i}$ are the Lagrange multipliers, and $\mu$ is a positive scalar. 

Under the ADMM framework, the approach is to iteratively update the three sets of variables $(\mathcal{X}_1, \mathcal{X}_2 ,\dots, \mathcal{X}_N),\mathcal{E}, (\mathcal{Y}_1, \mathcal{Y}_2, \dots, \mathcal{Y}_N)$. To be specific, at the start of the $k+1^\text{th}$ iteration, we fix $\mathcal{E} = \mathcal{E}^k$ and $\mathcal{Y}_\mathnormal{i} = \mathcal{Y}_\mathnormal{i}^k$, then for each $i$ solve:
\begin{equation}
\label{eq:min_x}
    \mathcal{X}_i^{k+1} = \underset{\mathcal{X}_i}{\text{argmin }} \mathcal{L}(\mathcal{X}_i,\mathcal{E}^k, \mathcal{Y}_i^k;\mu).
\end{equation}
Then, we fix $\mathcal{X}_i = \mathcal{X}_\mathnormal{i}^{k+1}$ and $\mathcal{Y}_\mathnormal{i} = \mathcal{Y}_\mathnormal{i}^k$ to solve:
\begin{equation}
\label{eq:min_e}
    \mathcal{E}^{k+1} = \underset{\mathcal{E}}{\text{argmin }} \mathcal{L}(\mathcal{X}_i^{k+1},\mathcal{E}, \mathcal{Y}_i^k;\mu).
\end{equation}
Finally we fix $\mathcal{X}_i = \mathcal{X}_\mathnormal{i}^{k+1}$ and $\mathcal{E} = \mathcal{E}^{k+1}$, and update $\mathcal{Y}_\mathnormal{i}^k$:
\begin{equation}
\label{eq:min_y}
    \mathcal{Y}_i^{k+1} = \mathcal{Y}_i^{k} + \mu(\mathcal{B} - \mathcal{X}_\mathnormal{i}^{k+1}- \mathcal{E}^{k+1}) .
\end{equation}

Next we derive closed form solutions for problem \eqref{eq:min_x} and for problem~\eqref{eq:min_e}. Problem \eqref{eq:min_x}, written out, reads:
\begin{equation}
\label{eq:Xi}
\begin{aligned}
& \mathcal{X}_i^{k+1} = \underset{\mathcal{X}_i}{\text{argmin }}
 \|\mathbf{X}_{i(i)}\|_*+   \frac{\mu}{2} \|\mathcal{X_\mathnormal{i}+E^\mathnormal{k}-B)}\|_F^2-\langle \mathcal{Y_\mathnormal{i}^\mathnormal{k},X_\mathnormal{i}+E^\mathnormal{k}-B } \rangle.
\end{aligned}
\end{equation}
Using the property of the Frobenius norm, $ \|\mathcal{A}_1+\mathcal{A}_2\|_F^2 = \|\mathcal{A}_1\|_F^2+ \|\mathcal{A}_2\|_F^2 + 2\langle\mathcal{A}_1,\mathcal{A}_2 \rangle $, problem \eqref{eq:Xi} can be written as:
\begin{equation}
\label{eq:Xi_new}
\begin{aligned}
 \mathcal{X}_i^{k+1} &= &&\underset{\mathcal{X}_i}{\text{argmin }}
 \|\mathbf{X}_{i(i)}\|_*+  \frac{\mu}{2} \left\lVert \frac{1}{\mu}\mathcal{Y_\mathnormal{i}^\mathnormal{k}+ \mathcal{B-E^\mathnormal{k}-X_\mathnormal{i} }} \right\rVert_F^2\\
 &=&& \underset{\mathcal{X}_i}{\text{argmin }}
 \|\mathbf{X}_{i(i)}\|_*+  \frac{\mu}{2} \left\lVert \frac{1}{\mu}\mathbf{Y}_{i(i)}^{k} + \mathbf{B}_{(i)}-\mathbf{E}_{(i)}^{k}-\mathbf{X}_\mathnormal{i(i)} \right\rVert_F^2.
\end{aligned}
\end{equation}
In the second line of~\eqref{eq:Xi_new}, we change the Frobenius norm of a tensor into the Frobenius norm of its $i$-th unfolding, which does not change the actual value of the norm. As a result, the objective function of problem~\eqref{eq:Xi_new} only involves matrices, so we can solve for $\mathbf{X}_{i(i)}$ using the  well-established closed form solution (e.g., see proof in Cai et. al~\cite{cai2010singular}):
$\mathbf{X}_{i(i)}^{k+1} = \mathbf{T}_{\frac{1}{\mu}}(\frac{1}{\mu}\mathbf{Y}_{i(i)}^{k} + \mathbf{B}_{(i)}-\mathbf{E}_{(i)}^{k}).$
Then we fold the matrix $\mathbf{X}_{i(i)}^{k+1}$ back into a tensor, i.e., $\mathcal{X}_i^{k+1} = \text{fold}_i(\mathbf{X}_{i(i)}^{k+1})$. The truncation operator $\mathbf{T}_\tau(\mathbf{X)}$ in for a matrix $\mathbf{X}= U\Sigma V^T$ is $\mathbf{T}_\tau(\mathbf{X)} = U\Sigma_{\Bar{\tau}}V^T,$
where  $\Sigma = \text{diag}(\sigma_i)$ is the eigenvalue diagonal matrix for $\mathbf{X}$. The operation $\Sigma_{\Bar{\tau}} = \text{diag(max}(\sigma_i-\tau, 0))$ discards the eigenvalues less than  $\tau$, and shrinks the remaining eigenvalues by $\tau$. 

We now proceed to derive a closed form solution to update $\mathcal{E}$ in problem~\eqref{eq:min_e}. Problem~\eqref{eq:min_e} is equivalent to solving:
\begin{equation}
\label{eq:E}
\mathcal{E}^{k+1}= \underset{\mathcal{E}}{\text{argmin }}
\lambda \|\mathbf{E}_{(1)}\|_{2,1} + \sum_{i=1}^N( \mu \|\mathcal{X}_{i}^{k+1}+\mathcal{E-B})\|_F^2-\langle \mathcal{Y}_{i}^{k},\mathcal{X}_{i}^{k+1}+\mathcal{E-B}  \rangle).
\end{equation}
Following the same technique as earlier, \eqref{eq:E} is equivalent to:
\begin{equation}
\label{eq:E_2}
\begin{aligned}
\mathcal{E}^{k+1}= & \underset{\mathcal{E}}{\text{argmin }}
 \lambda \|\mathbf{E}_{(1)}\|_{2,1}+ \sum_{i=1}^N\left( \frac{\mu}{2} \left\lVert \frac{1}{\mu}\mathcal{Y}_{i}^{k} + \mathcal{B}- \mathcal{X}_i^{k+1} - \mathcal{E} \right\rVert_F^2\right).
\end{aligned}
\end{equation}
By the proof of Goldfarb and Qin~\cite{goldfarb2014robust}, problem~\eqref{eq:E_2} shares the same solution with:
\begin{equation}\label{eq:E_3}
\begin{aligned}
\mathcal{E}^{k+1} = & \underset{\mathcal{E}}{\text{argmin}}~\lambda \left\lVert\mathbf{E}_{(1)}\right\rVert_{2,1}+ \frac{\mu N}{2} \left\lVert\mathcal{E} -\frac{1}{N}\sum_{i=1}^N \left( \frac{1}{\mu}\mathcal{Y}_{i}^{k} + \mathcal{B-X}_\mathnormal{i}^{k+1}\right)\right\rVert_F^2,
 \end{aligned}
\end{equation}
since they have the same first-order conditions. In order to simplify expression~\eqref{eq:E_3}, we denote the term $\frac{1}{N}\sum_{i=1}^N\left(\frac{1}{\mu}\mathcal{Y}_{i}^{k} + \mathcal{B-X}_\mathnormal{i}^{k+1}\right)$ by a single variable $\mathcal{C} \in \mathbb{R}^{I_1 \times I_2 \times\dots \times I_N}$. Thus,
\begin{equation}
\label{eq:E_4}
\begin{aligned}
\mathcal{E}^{k+1} &= & & \underset{\mathcal{E}}{\text{argmin }}
 \lambda \|\mathbf{E}_{(1)}\|_{2,1}+ \frac{\mu N}{2} \|  \mathcal{E - C} \|_F^2\\
 & = && \underset{\mathcal{E}}{\text{argmin }}
 \lambda \|\mathbf{E}_{(1)}\|_{2,1}+ \frac{\mu N}{2} \|  \mathbf{E}_{(1)} - \mathbf{C}_{(1)} \|_F^2,
 \end{aligned}
\end{equation}
where in the second line we use the same approach as in~\eqref{eq:Xi_new} in which we replace the tensor Frobenius norm by the equivalent Frobenius norm of the mode one unfolding. The objective function of problem~\eqref{eq:E_4} only involves matrices, and the closed form solution is~\cite{tang2011robust}:
\begin{equation}
\begin{aligned}
\mathbf{E}_{(1)j}^{k+1} = \mathbf{C}_{(1)j}~\text{max}\left\{0,1-\frac{\lambda}{\mu N\|\mathbf{C}_{(1)j}\|_2}\right\}, \text{ for } j = 1,2,\dots ,p,
 \end{aligned}
\end{equation}
where $\mathbf{E}_{(1)j}$ is the $j^{th}$ column of $\mathbf{E}_{(1)}$,  $\mathbf{C}_{(1)j}$ is the $j^{th}$ column of  $\mathbf{C}_{(1)}$, and the integer $p = I_2 \times I_3 \times \dots \times I_N$ is the total number of columns in $\mathbf{C}_{(1)}$. This operation effectively sets a column of  $\mathbf{C}_{(1)}$ to zero if its $l_2$ norm is less than $\frac{\lambda}{\mu N}$, and scales the elements down by a factor $1-\frac{\lambda}{\mu N\|\mathbf{C}_{(1)j}\|_2}$ otherwise~\cite{tang2011robust}.

Note that compared with the ADMM method where we just update $\mathcal{X}_i$ and $\mathcal{E}$ once, the  \textit{augmented Lagrangian multipliers} (ALM) method~\cite{lin2010augmented} seeks to find the exact solutions for primal variables   $\mathcal{X}_i$ and $\mathcal{E}$ before updating Lagrangian multipliers $\mathcal{Y}_i =\mathcal{Y}_i^k $, yielding the framework as
\begin{equation*}
\begin{aligned}  
    (\mathcal{X}_i^{k+1},\mathcal{E}^{k+1} ) =& \underset{\mathcal{X}_i,\mathcal{E}}{\text{argmin }} \mathcal{L}(\mathcal{X}_i,\mathcal{E}, \mathcal{Y}_i^k;\mu)\\
    \mathcal{Y}_i^{k+1} = &\mathcal{Y}_i^{k} + \mu(\mathcal{B} - \mathcal{X}_\mathnormal{i}^{k+1}- \mathcal{E}^{k+1}).
\end{aligned}
\end{equation*}
As pointed out by Lin et al.~\cite{lin2010augmented},  compared with ALM, not only is ADMM  still able to converge to the optimal solution for  $\mathcal{X}_i$ and $\mathcal{E}$, but also the speed performance is better.
It is also noted that while in ALM, the $\mathcal{X}$ and $\mathcal{E}$ are optimized jointly, in the ADMM implementation, they are in fact updated sequentially~\cite{boyd2011distributed}. It is often observed in the matrix settings (see e.g.,  Lin et al.~\cite{lin2010augmented}) that updating the term containing outliers before the low rank term (i.e., $\mathcal{E}$ before $\mathcal{X}$ in the tensor setting) the low rank term results in faster convergence. As a consequence this is the  approach followed in the numerical implementation of Algorithm \ref{alg:ADMM21} used later in this work. 

\begin{algorithm}[t]
  \caption{Tensor robust PCA for fiber-wise corruptions}\label{alg:ADMM21}
  \begin{algorithmic}[1]
     \State Given $\mathcal{B},\lambda, \mu$. Initialize $\mathcal{X}_i = \mathcal{E}=\mathcal{Y}_i=\mathbf{0}.$
     \For{$k = 0,1,\dots $}
        \For{$i = 1:N$} \Comment{Update $\mathcal{X}$  }
        \State $\mathbf{X}_{i(i)}^{k+1} = \mathbf{T}_{\frac{1}{\mu}}\left(\frac{1}{\mu}\mathbf{Y}_{i(i)}^{k} + \mathbf{B}_{(i)}-\mathbf{E}_{(i)}^k\right).$
        \State $\mathcal{X}_i^{k+1} = \text{fold}_i\left(\mathbf{X}_{i(i)}^{k+1}\right)$
        \EndFor
        \State $\mathcal{C} = \frac{1}{N}\sum_{i=1}^N\left( \frac{1}{\mu}\mathcal{Y}_{i}^{k+1} + \mathcal{B}-\mathcal{X}_{i}^{k+1}\right)$ \Comment{Update $\mathcal{E}.$ }
        \For{$j = 1,2, \dots, p$}
        \State  $\mathbf{E}_{(1)j}^{k+1} = \mathbf{C}_{(1)j}\text{max}\left\{0,1-\frac{\lambda}{\mu N\|\mathbf{C}_{(1)j}\|_2}\right\}$
        \EndFor
        \State  $\mathcal{E}^{k+1} = \text{fold}_1(\mathbf{E}_{(1)}^{k+1})$
        \For{$i = 1:N$} \Comment{Update $\mathcal{Y}.$  }
        \State $ \mathcal{Y}_i^{k+1} = \mathcal{Y}_i^{k} + \mu(\mathcal{B} - \mathcal{X}_\mathnormal{i}^{k+1}- \mathcal{E}^{k+1}) . $
    \EndFor
    \EndFor
    \State \Return $\mathcal{X}^{k} = \frac{1}{N}\sum_{i=1}^N\mathcal{X}_i^{k}, \mathcal{E}^{k}$
  \end{algorithmic}
\end{algorithm}

In the implementation of Algorithm~\ref{alg:ADMM21}, we set the convergence criterion as
\begin{equation} \label{eq:converge}
    \frac{\|\mathcal{B-E-X}\|_F}{\|\mathcal{B}\|_F} \leq \epsilon,
\end{equation}
Where $\epsilon$ is the tolerance. 


\subsubsection{Partial Observation}\label{partial observation}
Now we provide an algorithm to solve problem \eqref{eq:TC}. Similar to the matrix setting in Tang et al.~\cite{tang2011robust}, we set the fibers of the low rank tensor to be zero in the locations corresponding to outliers. We introduce a compensation tensor $\mathcal{O}\in \mathbb{R}^{I_1 \times I_2 \times\dots \times I_N}$, which is zero for entries in the observation set $\Omega$, and can take any value outside $\Omega$. Thus using the same auxiliary variables technique as in~\eqref{eq:HoPCA-C-xn}, we can reformulate problem \eqref{eq:TC} as:
\begin{equation}
\label{eq:TC2}
\begin{aligned}
& \underset{\mathcal{X}_i,\mathcal{E}}{\text{min}}
& & \sum_{i=1}^N\|\mathbf{X}_{i(i)}\|_*+ \lambda \|\mathbf{E}_{(1)}\|_{2,1}\\ 
& \text{s.t.}
& & \mathcal{B = X_\mathnormal{i}+E+O}, i = 1,2,\dots,\ N,\\&
& & \mathcal{O}_\Omega = 0.
\end{aligned}
\end{equation}
Since $\mathcal{O}$ compensates for whatever the value is in the unobserved entries of $\mathcal{B}$, we only need to keep track of the indices of the unobserved entries, and can simply set the unobserved entries of $\mathcal{B}$ to zero. The augmented Lagrangian function for problem \eqref{eq:TC2} is 
\begin{equation*}
\begin{aligned}
    \mathcal{L}(\mathcal{X}_1, \mathcal{X}_2, \dots, \mathcal{X}_N,\mathcal{E},\mathcal{O}, \mathcal{Y}_1, \mathcal{Y}_2, \dots, \mathcal{Y}_N;\mu)=
    \sum_{i=1}^N\|\mathbf{X}_{i(i)}\left\lVert _*+ \lambda\right\rVert \mathbf{E}_{(1)}\|_{2,1}  + \\ \sum_{i=1}^N\left( \frac{\mu}{2} \|\mathcal{X_\mathnormal{i}+E+O-B)}\|_F^2-\langle \mathcal{Y_\mathnormal{i},X_\mathnormal{i}+E+O-B }\rangle\right).
\end{aligned}
\end{equation*}

We again use the ADMM framework now iteratively updating $\mathcal{X}_i$, $\mathcal{E}$,  $\mathcal{O}$ and $\mathcal{Y}_i$. The proof of the closed form solution for updating $\mathcal{X}_i$, $\mathcal{E}$ and $\mathcal{Y}_i$ is similar to Algorithm \ref{alg:ADMM21}. For $\mathcal{O}$, we fix $\mathcal{X}_i = \mathcal{X}_\mathnormal{i}^{k+1}$, $\mathcal{E} = \mathcal{E}^{k+1}$ and $\mathcal{Y}_\mathnormal{i} = \mathcal{Y}_\mathnormal{i}^k$, to solve \eqref{eq:min_o}:
\begin{equation} \label{eq:min_o}
\begin{aligned}
    \mathcal{O}^{k+1} &= &\underset{\mathcal{O}}{\text{argmin}}&\quad \mathcal{L}(\mathcal{X}_i^{k+1},\mathcal{E}^{k+1},\mathcal{O}, \mathcal{Y}_i^k;\mu)\\
     &&\text{s.t.} & \quad \mathcal{O}_\Omega = 0.
\end{aligned}
\end{equation}

Following the same procedure as before (see \eqref{eq:E}, \eqref{eq:E_2} and \eqref{eq:E_3}), we have:

\begin{equation} \label{eq:o}
\begin{aligned}
 \mathcal{O}^{k+1} &=  &\underset{\mathcal{O}}{\text{argmin }} &\quad
    \sum_{i=1}^N\left(  \frac{\mu}{2} \left\lVert\mathcal{X}_{i}^{k+1}+\mathcal{E}^{k+1}+\mathcal{O}-\mathcal{B})\right\rVert_F^2-\langle \mathcal{Y}_{i}^{k},\mathcal{X}_{i}^{k+1}+\mathcal{E}^{k+1}+\mathcal{O}-\mathcal{B}  \rangle\right)\\
    &= & \underset{\mathcal{O}}{\text{argmin }} & \quad  \sum_{i=1}^N\left(  \frac{\mu}{2} \left\lVert   \frac{1}{\mu}\mathcal{Y}_i^k +\mathcal{B}-\mathcal{X}_{i}^{k+1}- \mathcal{E}^{k+1}-\mathcal{O} \right\rVert_F^2\right)\\
    & =& \underset{\mathcal{O}}{\text{argmin }} & \quad  \frac{\mu N}{2} \left\lVert \mathcal{O} - \frac{1}{N}\sum_{i=1}^{N} \left(\frac{1}{\mu}\mathcal{Y}_i^k +\mathcal{B-X}_i^{k+1} - \mathcal{E}^{k+1}\right) \right\rVert_F^2,\\
    & & \text{s.t.} & \quad \mathcal{O}_\Omega = 0.
\end{aligned}
\end{equation}
For \eqref{eq:o}, we simply set $\mathcal{O}= \frac{1}{N}\sum_{i=1}^{N} (\frac{1}{\mu}\mathcal{Y}_i^k +\mathcal{B-X}_i^{k+1} - \mathcal{E}^{k+1})$ for entries $(I_1, I_2, \dots, I_N) \in \Omega^C$, and zero otherwise. The procedure is summarized in Algorithm~\ref{alg:TC}.
\begin{algorithm}[t]
  \caption{ADMM for robust tensor completion}\label{alg:TC}
  \begin{algorithmic}[1]
     \State Given $\mathcal{B},\lambda, \mu$. Initialize $\mathcal{X}_i = \mathcal{E}=\mathcal{Y}_i=\mathcal{O} = \mathbf{0}.$
     \For{$k$ = 0,1, $\cdots$ }
        \For{$i$ = 1:N} \Comment{Update $\mathcal{X}$  }
        \State $\mathbf{X}_{i(i)}^{k+1} = \mathbf{T}_{\frac{1}{\mu}}(\mathbf{B}_{(i)}+\frac{1}{\mu}\mathbf{Y}_{i(i)}^{k}-\mathbf{E}_{(i)}^k-\mathbf{O}^{k}_{(i)}),$
        \State $\mathcal{X}_i^{k+1} = \text{fold}_i(\mathbf{X}_{i(i)}^{k+1})$
        \EndFor
    \State $\mathcal{C} = \frac{1}{N}\sum_{i=1}^N\left( \frac{1}{\mu}\mathcal{Y}_i^{k+1} + \mathcal{B}-\mathcal{X}_{i}^{k+1}-\mathcal{O}^{k+1}\right)$ \Comment{Update $\mathcal{E}.$ }
        \For{$j = 1,2 ,\dots, p$}
    \State  $\mathbf{E}_{(1)j}^{k+1} = \mathbf{C}_{(1)j}\text{max}\left\{0,1-\frac{\lambda}{\mu N\|\mathbf{C}_{(1)j}\|_2}\right\}$
    \EndFor
    \State  $\mathcal{E}^{k+1} = \text{fold}_1(\mathbf{E}_{(1)}^{k+1})$
    \State $\mathcal{O}^{k+1} = \left(\sum_{i=i}^N\left(\frac{1}{\mu}\mathcal{Y}_i^{k} + \mathcal{B}-\mathcal{X}_i^{k+1}-\mathcal{E}^{k+1}\right)\right)_{{\Omega}^C}.$\label{lst:line:O}
    \Comment{Update $\mathcal{O}.$} 
    \For{$i = 1:N$} \Comment{Update $\mathcal{Y}.$  }
        \State $\mathcal{Y}_i^{k+1} =  \mathcal{Y}_i^{k} + \mu(\mathcal{B} - \mathcal{X}_\mathnormal{i}^{k+1}- \mathcal{E}^{k+1} - \mathcal{O}^{k+1}) $
    \EndFor
    \EndFor
    \State \Return $\mathcal{X}^{k}= \frac{1}{N}\sum_{i=1}^N\mathcal{X}_i^{k}, \mathcal{E}^{k}$
  \end{algorithmic}
\end{algorithm}

We set the convergence criterion of Algorithm~\ref{alg:TC} as
\begin{equation*}
    \frac{\|\mathcal{B-E-X-O}\|_F}{\|\mathcal{B}\|_F} \leq \epsilon,
\end{equation*}
which is similar to \eqref{eq:converge} but accounts for the additional tensor $\mathcal{O}$.

\section{Numerical experiments} \label{Sec:Simulation}

In this section, we apply Algorithms~\ref{alg:ADMM21} and~\ref{alg:TC} developed in Section \ref{Sec:Algorithm} on a series of test problems using synthetically generated datasets. We first conduct tensor decomposition on fiber-wise corrupted data, then we examine the case when the data is only partially observed and is also fiber-wise corrupted. For both cases, we compare the performance of our algorithms, which are $l_{2,1}$ norm constrained decomposition, with $l_1$ norm constrained decomposition~\cite{goldfarb2014robust, tan2013traffic}. We demonstrate via the numerical experiments that our algorithms are able to exactly recover the original low-rank tensor, and identify the position of corrupted fibers. In comparison, $l_1$ norm constrained algorithms performs poorly in the fiber-wise corrupted settings, unable to achieve exact recoveries.

\subsection{Performance measures and implementation}
For each of the numerical experiments, the performance of the algorithms are measured by the relative error of the low rank tensor, as well as the precision and recall of the outlier fibers. The \textit{relative error} (RE) of low rank tensor is calculated as:
\begin{equation*}
   \text{RE} = \frac{\|\mathcal{X}_0-\hat{\mathcal{X}}\|_F}{\|\mathcal{X}_0\|_F},
\end{equation*}
where $\mathcal{X}_0$ is the true low rank tensor modified to take the value 0 in the entries corresponding to the  corrupted fibers; $\hat{\mathcal{X}}$ is the estimated low rank tensor resulting from application of Algorithm~\ref{alg:ADMM21} or \ref{alg:TC}, which also has the value 0 in the fibers that are estimated to be corrupted.

We compute the \textit{precision} of the algorithm to assess the potential to correctly identify only the outlier fibers. It is computed as:
\begin{equation*}
  \text{precision}  = \frac{\text{tp}}{\text{tp} + \text{fp}},
\end{equation*}
where the \textit{true positives} (tp) corresponds the number of estimated outlier fibers which are true outliers, and the \textit{false positives} (fp) corresponds to the number of estimated outlier fibers which are not true outliers. 

The \textit{recall}, which measures the ability to find all outlier fibers, is defined as
\begin{equation*}
  \text{recall}  = \frac{\text{tp}}{\text{tp} + \text{fn}},
\end{equation*}
where the \textit{false negatives} (fn) correspond to the number of true outlier fibers that were not  correctly identified by the estimator.

For the convergence criterion we set $\epsilon = 10^{-7}$, and we use an empirical value $\lambda = \frac{1}{0.03I_m}$, where $I_m = \text{max}(I_1, \dots , I_N)$. The hyperparameter $\lambda$ in $l_1$ norm constrained decomposition algorithm is also tuned for its best performance in our settings.

All of the experiments are carried out on a Macbook Pro with quad-core 2.7GHz  Intel i7 Processor and 16GB RAM, running Matlab R2018a. We modify and extend the code of Lin et. al~\cite{lin2010augmented}, using PROPACK toolbox to efficiently calculate the SVD. The code is modified to update variables in line with the distinct problem formulation using the $l_{2,1}$ norm and to scale to tensors rather than matrices. The Tensor Toolbox for Matlab~\cite{TTB_Software}~\cite{TTB_Dense} is also used for tensor manipulations. The resulting source code is available at \url{https://github.com/Lab-Work/Robust_tensor_recovery_for_traffic_events}.

\subsection{Tensor robust PCA} \label{higher-order RPCA}
In this subsection we apply higher-order RPCA to the problems where we have fully observed data with fiber-wise corrupted entries. 

\subsubsection{Simulation conditions} We synthetically generate the observation data as $\mathcal{B} = \mathcal{X}_0 + \mathcal{E}_0 \in  \mathbb{R}^{I_1 \times I_2 \times I_3}$, where  $\mathcal{X}_0$ and $\mathcal{E}_0$ are the \textit{true} or ``ground truth'' low-rank tensor and fiber-sparse tensor, respectively. We generate $\mathcal{X}_0 \in  \mathbb{R}^{I_1 \times I_2 \times I_3}$ as a core tensor $\mathcal{G} \in \mathbb{R}^{c_1 \times c_2 \times c_3}$ with size $c_1\times c_2\times c_3$ and tucker rank $(c_1,c_2,c_3)$, multiplied in each mode by orthogonal matrices of corresponding dimensions, $\mathbf{U}^{(i)} \in \mathbb{R}^{I_i \times c_i}$:
\begin{equation*}
      \mathcal{X}_0= \mathcal{G} \times_1 \mathbf{U}^{(1)}\times_2 \mathbf{U}^{(2)} \times_3 \mathbf{U}^{(3)}.
\end{equation*}
The entries of $\mathcal{G}$ are independently sampled from standard Gaussian distribution. The orthogonal matrices $\mathbf{U}^{(i)}$ are generated via a Gram-Schmidt orthogonalization on $c_i$ vectors of size $\mathbb{R}^{I_i}$ drawn from standard Gaussian distribution. The sparse tensor $\mathcal{E}_0  \in  \mathbb{R}^{I_1 \times I_2 \times I_3}$ is formed by first generating a tensor $\mathcal{E}_0' \in \mathbb{R}^{I_1 \times I_2 \times I_3}$, whose entries are i.i.d uniform distribution $\mathcal{U}$(0,1). Then  we randomly keep a fraction $\gamma$ of the fibers of $\mathcal{E}_0'$ to form $\mathcal{E}_0$. Finally, the corresponding fibers of $\mathcal{X}_0$ with respect to non-zero fibers in $\mathcal{E}_0$ are set to zero.

\subsubsection{Algorithm performance for varying problem sizes} We apply Algorithm \ref{alg:ADMM21} on $\mathcal{B}$ of varying tensor sizes $(I_1, I_2, I_3)$ and underlying tucker rank $(c_1,c_2,c_3)$, and predict $\hat{\mathcal{X}}$ and $\hat{\mathcal{E}}$ using Algorithm \ref{alg:ADMM21}. We also apply $l_1$ norm constrained decomposition on the same settings. Table \ref{T-RPCA} compares the result. The corruption rate is set to $5\%$, i.e.,  $\gamma = 0.05$. In all cases, for our algorithm the relative residual errors are less than $10^{-6}$, which is the same precision that we set for convergence tolerance. That is to say, we can exactly recover the low rank tensors in this setting. The precision and recall are both 1.0, indicating that the outlier detection is also exact. Similar to the observation of Cand{\`e}s' et al.~\cite{candes2011robust}, the iteration numbers tend to be constant (between 28 and 29 in this case) regardless of tensor size. This indicates that the number of of SVD computations might be limited and insensitive to the size, which is important since SVD is the computational bottleneck of the algorithm.  Furthermore, this property is important to allow the problem to solve quickly even on datasets of moderate sizes, as will be shown in a case study in Section~\ref{Sec:case}. 

In comparison, although $l_1$ norm constrained decomposition can also detect outliers in high precision and recall, the relative residual errors are relatively, high, in the order of $10^{-1}$. This indicates that $l_1$ norm constrained decomposition can do an adequate job when corruption ratio is low ($\gamma = 0.05$), but cannot achieve exact recoveries. 

\begin{table}[t]
  \begin{center}
    \caption{Application of Algorithm \ref{alg:ADMM21} on fiber-wise corrupted tensors with full observation, and comparison with $l_1$ norm constrained decomposition. For different tensor sizes $(I_1,I_2,I_3)$ and Tucker ranks $(c_1,c_2,c_2)$, where we set $c = 0.1 I$, we show the relative errors of low rank tensors (RE), the precision and recall of outlier fibers identification, as well as the number of iterations (iter) and total time for convergence.}
    \label{T-RPCA}
    
    \subfloat[Algorithm~\ref{alg:ADMM21}: $l_{2,1}$ norm constrained decomposition]{
    \begin{tabular}{ccccccc} 
    \toprule
      $(I_1,I_2,I_3)$ & $(c_1,c_2,c_2)$ & RE & precision & recall & iter & time(s) \\
      \midrule
      (70,70,70) & (7,7,7) & $1.23\times 10^{-7}$ & 1.0 & 1.0 & 29 & 9.9\\
      (90,90,90) & (9,9,9) & $1.24\times 10^{-7}$ & 1.0 & 1.0 & 28 & 16.2 \\
      (150,150,150) & (15,15,15) & $6.68\times 10^{-8}$ & 1.0 & 1.0 & 28 & 50.5 \\
      (210,210,210) & (21,21,21) & $7.35\times 10^{-8}$ & 1.0 & 1.0 & 28 &  133.0 \\
      \bottomrule
    \end{tabular}}
    
    \subfloat[Comparison: $l_1$ norm constrained decomposition]{
    \begin{tabular}{ccccccc} 
    \toprule
      $(I_1,I_2,I_3)$ & $(c_1,c_2,c_2)$ & RE & precision & recall & iter & time(s) \\
      \midrule
      (70,70,70) & (7,7,7) & $2.10\times 10^{-1}$ & 1.0 & 1.0 & 28 & 1.4\\
      (90,90,90) & (9,9,9) & $2.28\times 10^{-1}$ & 1.0 & 1.0 & 29 & 2.6 \\
      (150,150,150) & (15,15,15) & $2.23\times 10^{-1}$ & 1.0 & 1.0 & 28 & 14.7 \\
      (210,210,210) & (21,21,21) & $2.27\times 10^{-1}$ & 0.99 & 1.0 & 35 &  101.0 \\
      \bottomrule
    \end{tabular}}
  \end{center}
\end{table}

\subsubsection{Influence of the corruption rate} 
\begin{figure}
\centering
\subfloat[]{\includegraphics[width=0.45\textwidth]{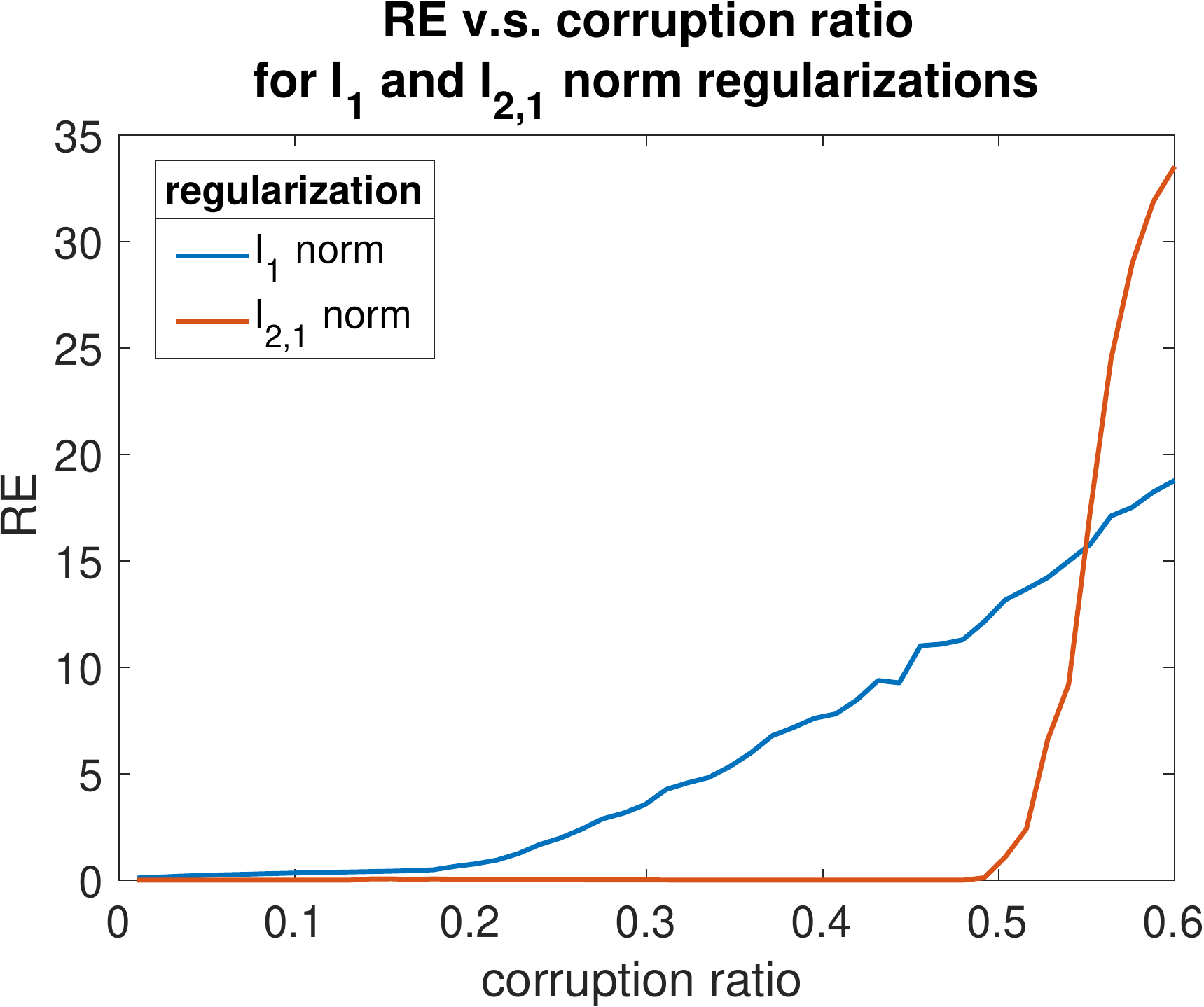}
\label{fig:errL_HoRPCA}}
\hfil
\subfloat[]{\includegraphics[width=0.45\textwidth]{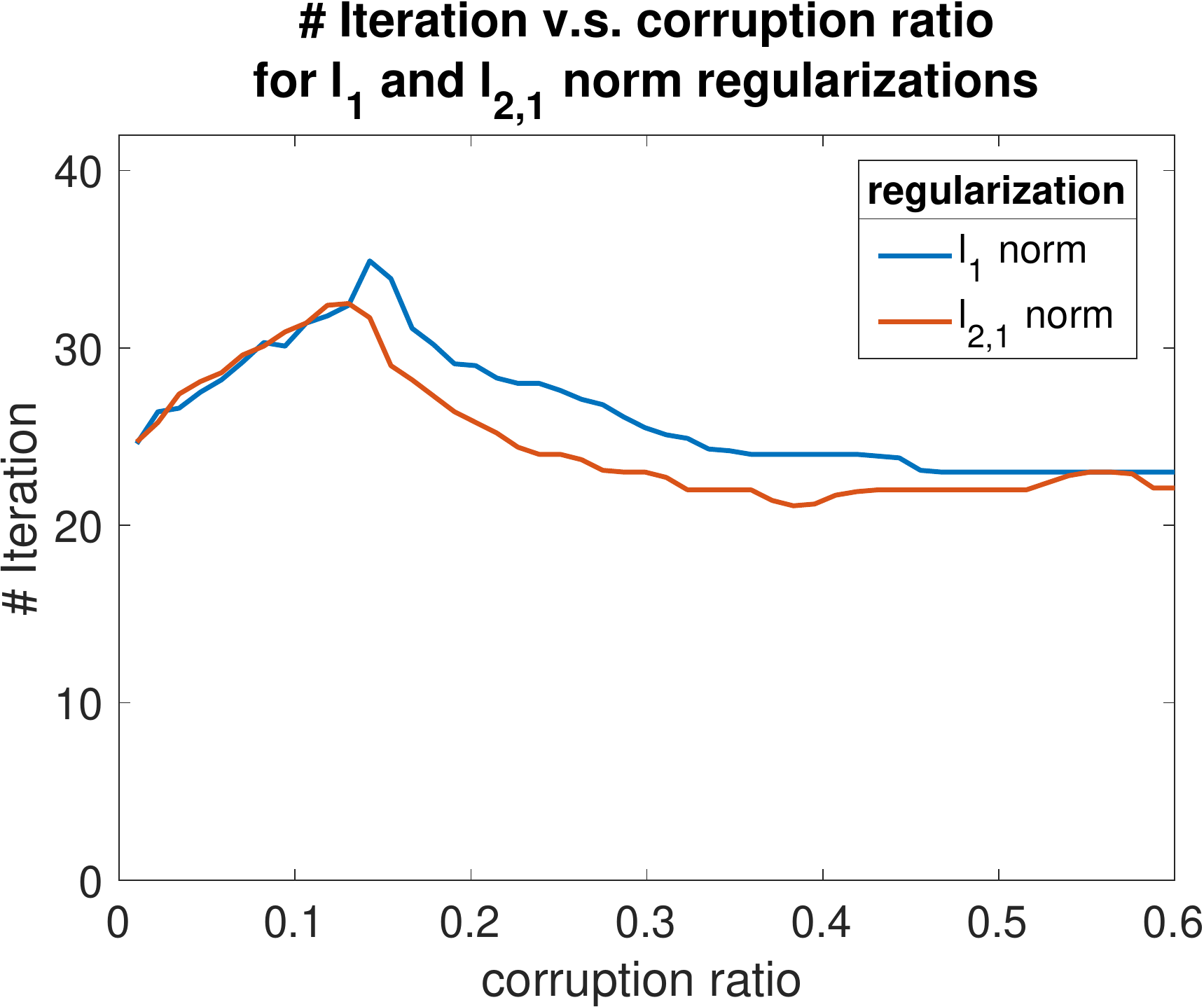}
\label{fig:iter_HoRPCA}}
\hfil
\subfloat[]{\includegraphics[width=0.45\textwidth]{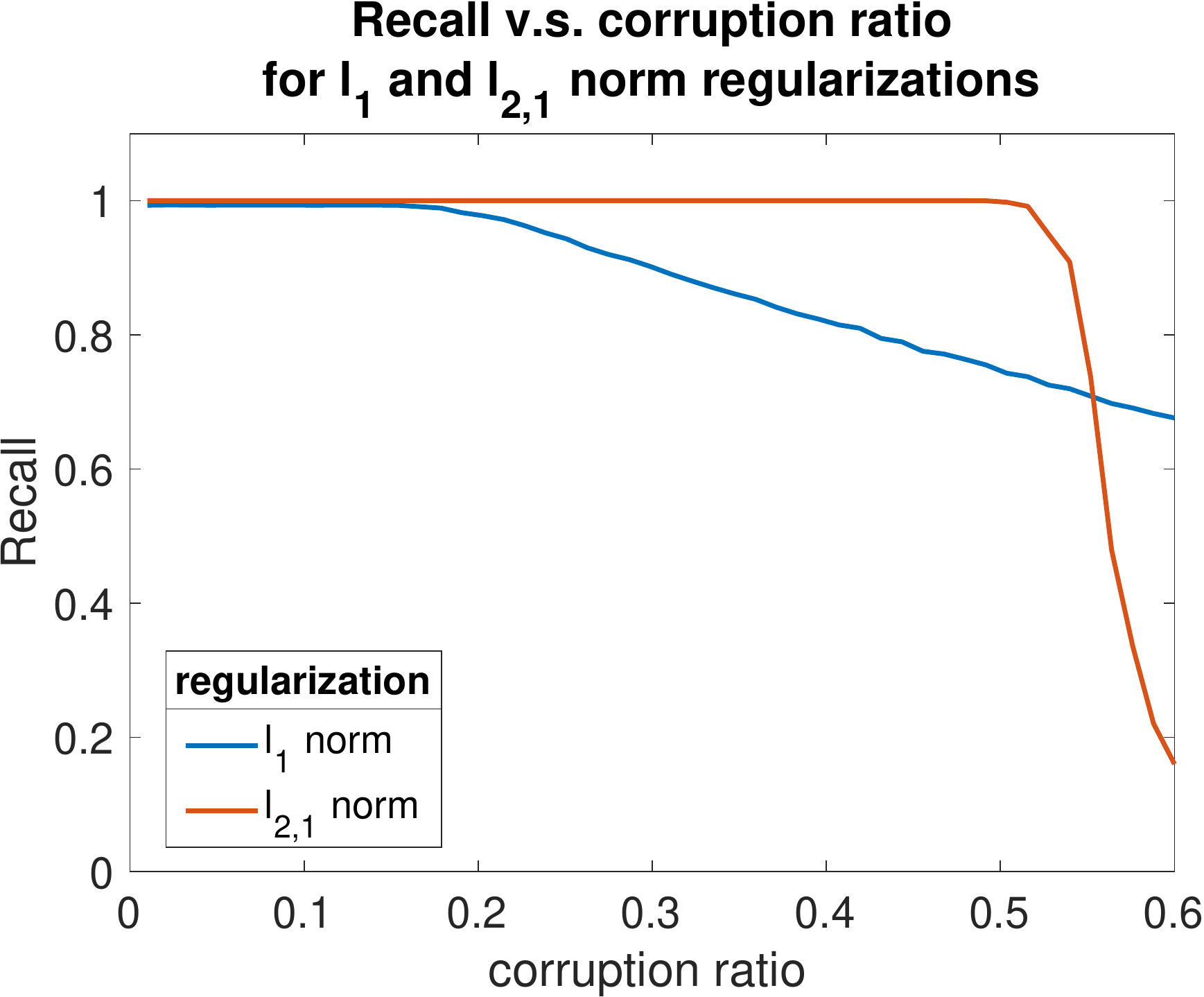}
\label{fig:recall_HoRPCA}}
\hfil
\subfloat[]{\includegraphics[width=0.45\textwidth]{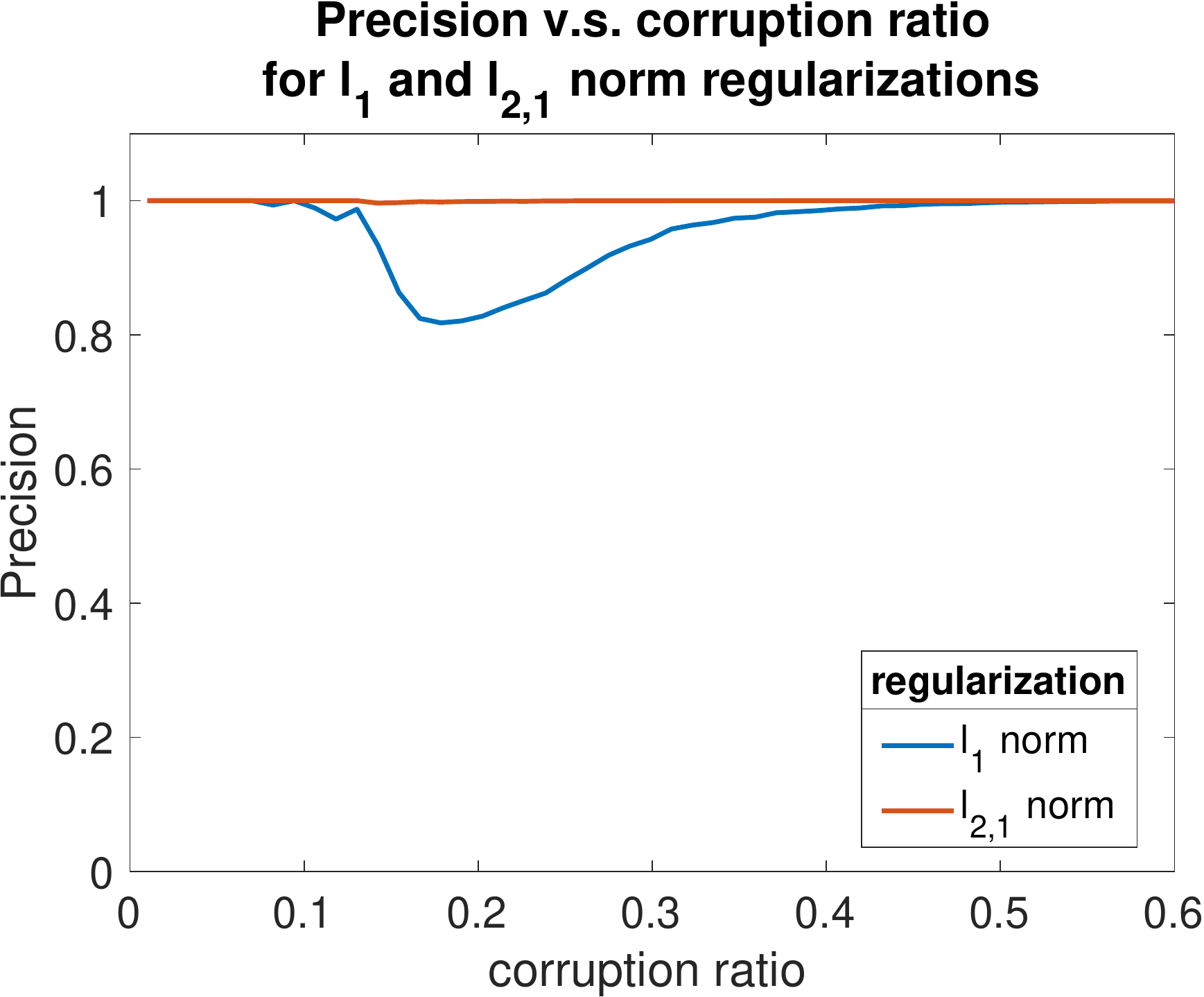}
\label{fig:precision_HoRPCA}}
\caption{Comparison of Algorithm~\ref{alg:ADMM21} ($l_{2,1}$ norm regularized tensor decomposition), with $l_1$ norm regularized tensor decomposition, as gross corruption rate changes. We fix the tensor size at $\mathbb{R}^{70 \times 70 \times 70}$, and fix the tucker rank of low rank tensor $\mathcal{X}_0$ at $(5,5,5)$, then vary the gross corruption ratio $\gamma$ from 0\% to 100\%. The result is an average over 10 trials.}
\label{fig:HoRPCA}
\end{figure}

Next, we investigate the performance of Algorithm \ref{alg:ADMM21} as the corruption ratio changes, and compare the result with $l_1$ norm constrained decomposition. We fix the low-rank tensor $\mathcal{X}_0$ at size $  \mathbb{R}^{70 \times 70 \times 70}$ with a tucker rank of $(5,5,5)$, then vary the gross corruption ratio $\gamma$ from 0\% to 60\%. The results are shown in Figure \ref{fig:HoRPCA} as an average over 10 trials. In this setting we see that as long as the corruption ratio is below 0.47, Algorithm~\ref{alg:ADMM21} can precisely recover the low rank tensor, and correctly identify the outlier fibers. On the other hand, the relative error of $l_1$ norm constrained decomposition is constantly higher. After the corruption ratio exceeds 0.2, the estimation is no longer useful, with the relative error exceeding 100\%.

\subsection{Robust tensor completion} \label{subsec:TC_sim}
In this subsection we look at the performance of Algorithm \ref{alg:TC}, when the data is only partially observed, and compare it with $l_1$ norm constrained decomposition.

\subsubsection{Simulation conditions} We first generate a full observation data $\mathcal{B}' = \mathcal{X}_0 + \mathcal{E}_0 \in  \mathbb{R}^{I_1 \times I_2 \times I_3}$ in the same way as Section \ref{higher-order RPCA}. $\mathcal{X}_0 \in \mathbb{R}^{I_1 \times I_2 \times I_3}$ is the low rank tensor, and  $\mathcal{E}_0 \in  \mathbb{R}^{I_1 \times I_2 \times I_3}$ is the sparse tensor. Then, we form the partial observation data $\mathcal{B} \in  \mathbb{R}^{I_1 \times I_2 \times I_3}$ by randomly keeping a fraction $\rho$ of the entries in $\mathcal{B}'$. We record the indices of the unobserved entries, and set their values in $\mathcal{B}$ as 0.

\subsubsection{Influence of the corruption and observation ratios} First, we apply Algorithm \ref{alg:TC} on simulated data with a varying corruption ratio and observation ratio. We fix the low-rank tensor $\mathcal{X}_0$ at size $\mathbb{R}^{70 \times 70 \times 70}$ with a tucker rank of $(5,5,5)$. For gross corruption ratio  $\gamma$ at 0.05, 0.1, 0.2, we vary the observation ratio $\rho$ from 0.1 to 1 and run Algorithm \ref{alg:TC}. We run 10 times and average the results.

\begin{figure}
\centering
\subfloat[]{\includegraphics[width=0.45\textwidth]{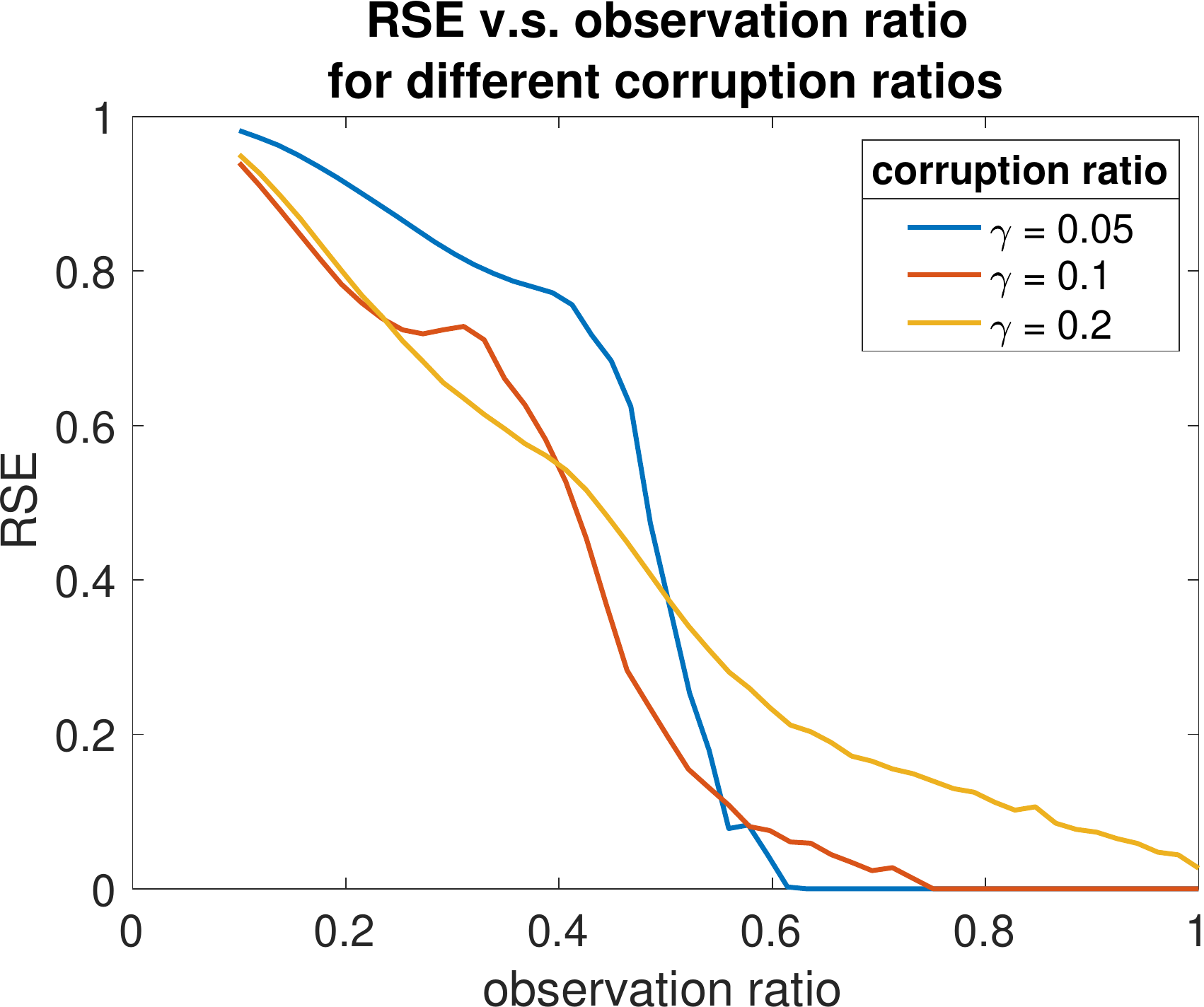}
\label{fig:TC_cor_errL}}
\hfil
\subfloat[]{\includegraphics[width=0.45\textwidth]{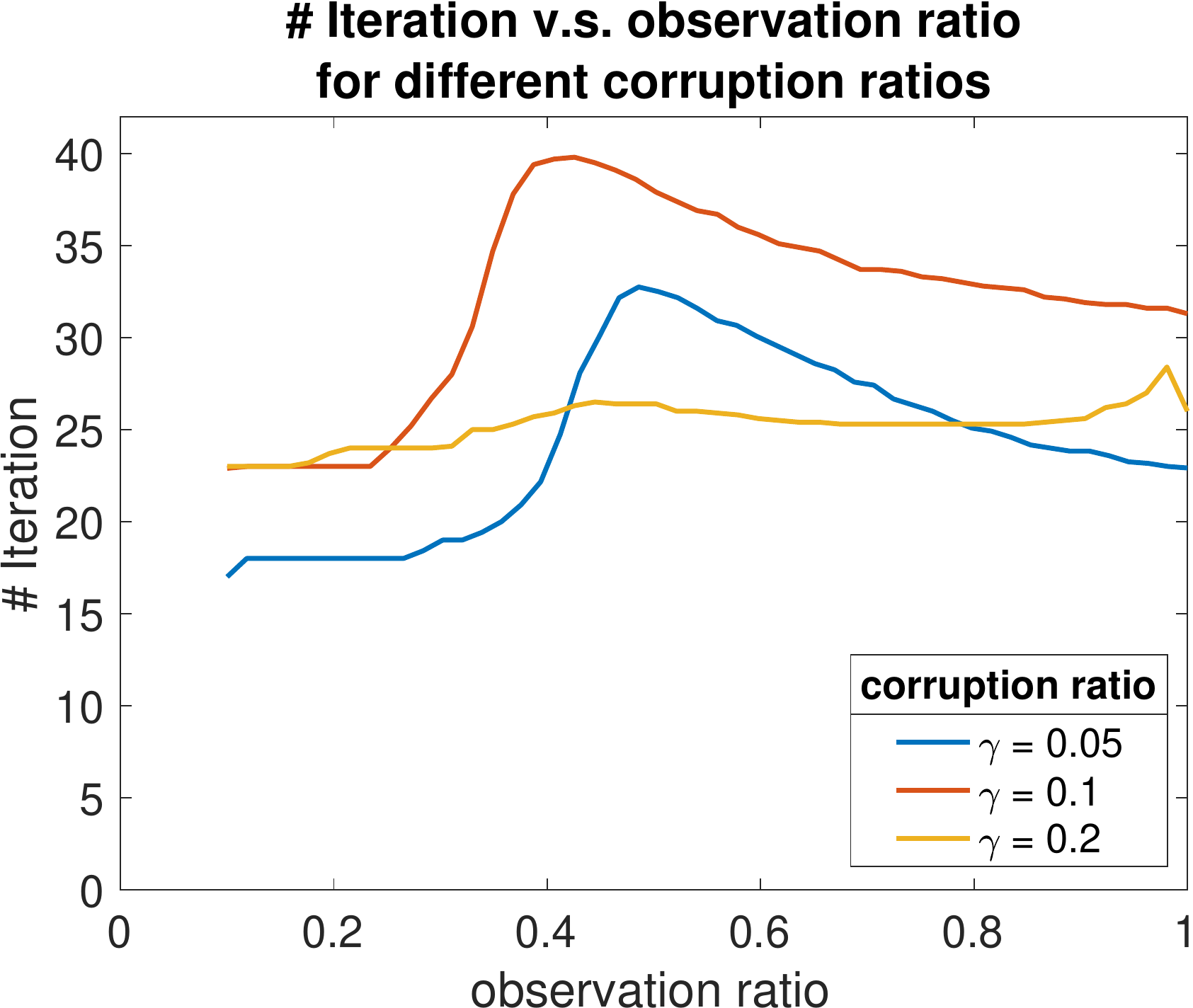}
\label{fig:TC_cor_iter}}
\hfil
\subfloat[]{\includegraphics[width=0.45\textwidth]{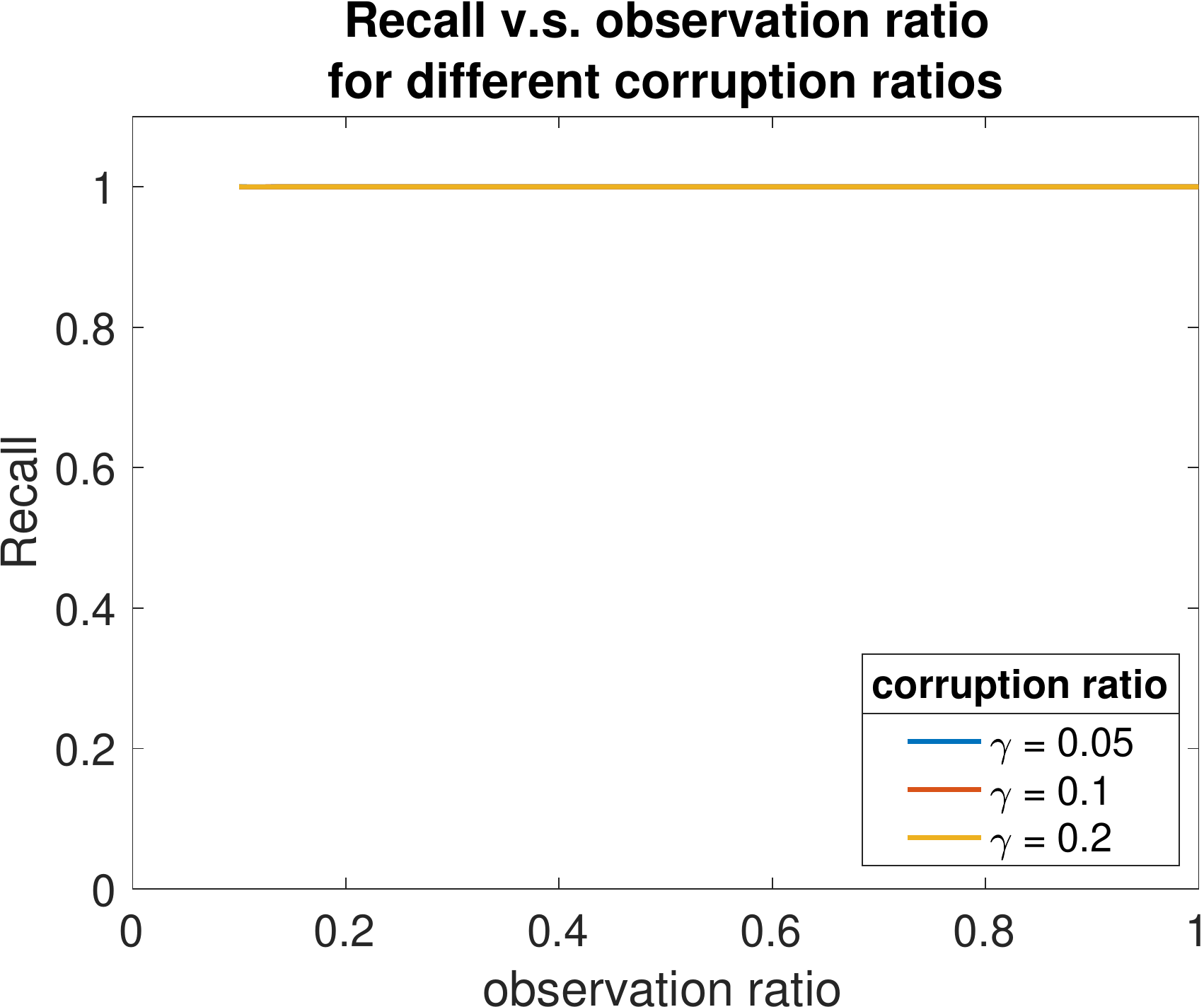}
\label{fig:TC_cor_recall}}
\hfil
\subfloat[]{\includegraphics[width=0.45\textwidth]{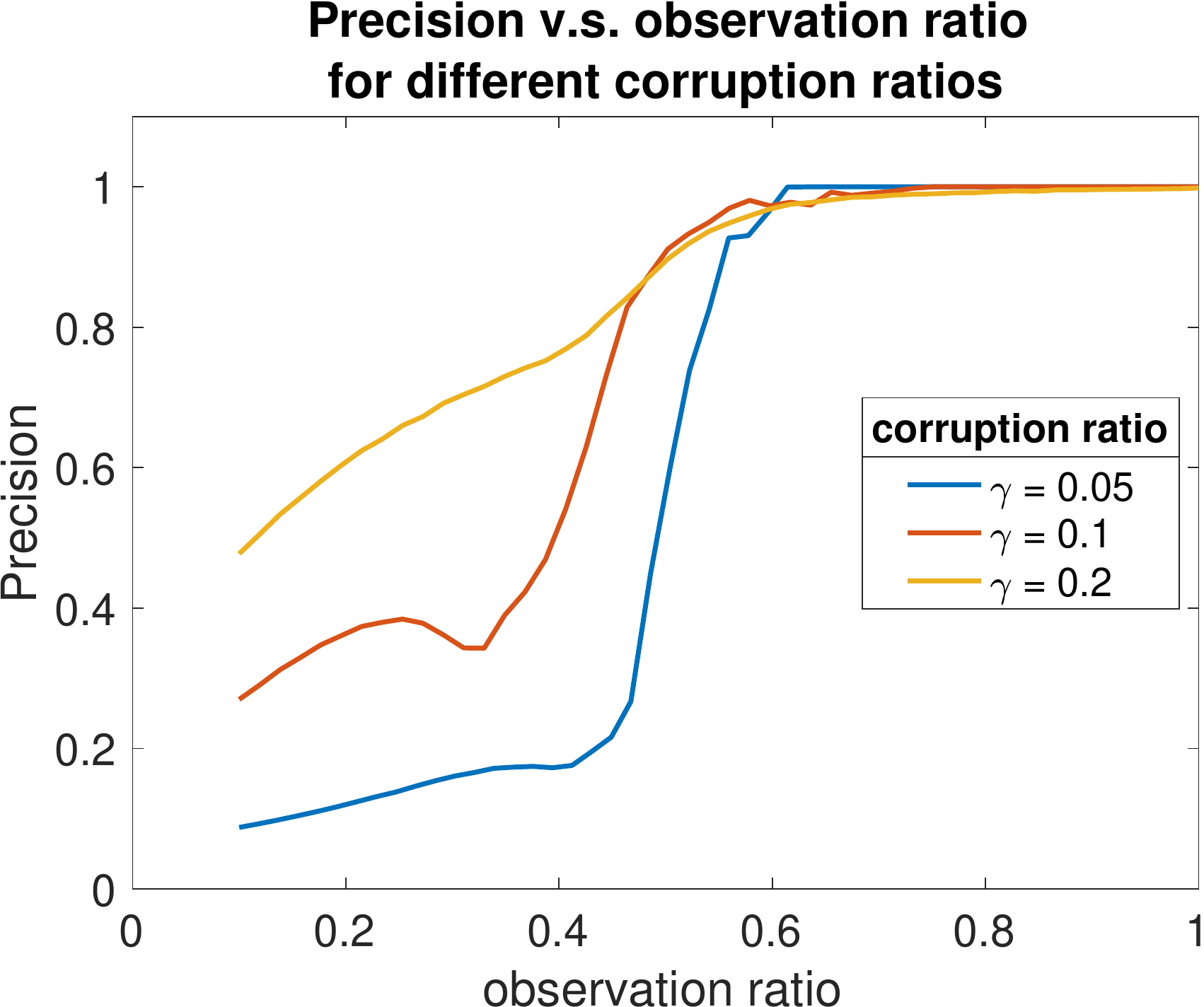}
\label{fig:TC_cor_precision}}
\caption{Results of Algorithm~\ref{alg:ADMM21} as a function of observation ratio with different corruption rates. The low-rank tensor $\mathcal{X}_0$ size is fixed at $\mathbb{R}^{70 \times 70 \times 70}$ with a tucker rank of $(5,5,5)$, and the gross corruption ratio $\gamma$ is set at 0.05, 0.1, or 0.2. The result is an average over 10 trials.}
\label{fig:TC_cor}
\end{figure}

Figure \ref{fig:TC_cor} shows that for the detection of corrupted fibers, the recall stays at 1. The precision stays at 1 when $\rho$ is above 0.6 but drops dramatically for smaller $\rho$. The relative error of low rank tensor is zero when the observation ratio $\rho > 0.6$ with corruption ratio  $\gamma = 0.05$, and when the observation ratio $\rho > 0.8$ with $\gamma = 0.1$. We observe a phase-transition behavior, in that the decomposition is exact when the observation ratio $\rho$ is above a critical threshold, but the performance drops dramatically below the threshold. This critical threshold on the observation ratio $\rho$ varies for each case, namely the algorithm can handle more missing entries as the number of outliers $\gamma$ is reduced. But when the corruption ratio is too large, exact recovery is not guaranteed. 

Overall, the performance is promising, since we can exactly identify the outlier positions and recover the low rank tensor at non-corrupted entries, even with relatively a large missing ratio, for example when 5\% of the fibers are corrupted and 40\% of the data is missing.

\subsubsection{Influence of the observation ratio and the tensor rank} 
\begin{figure}
\centering
\subfloat[]{\includegraphics[width=0.45\textwidth]{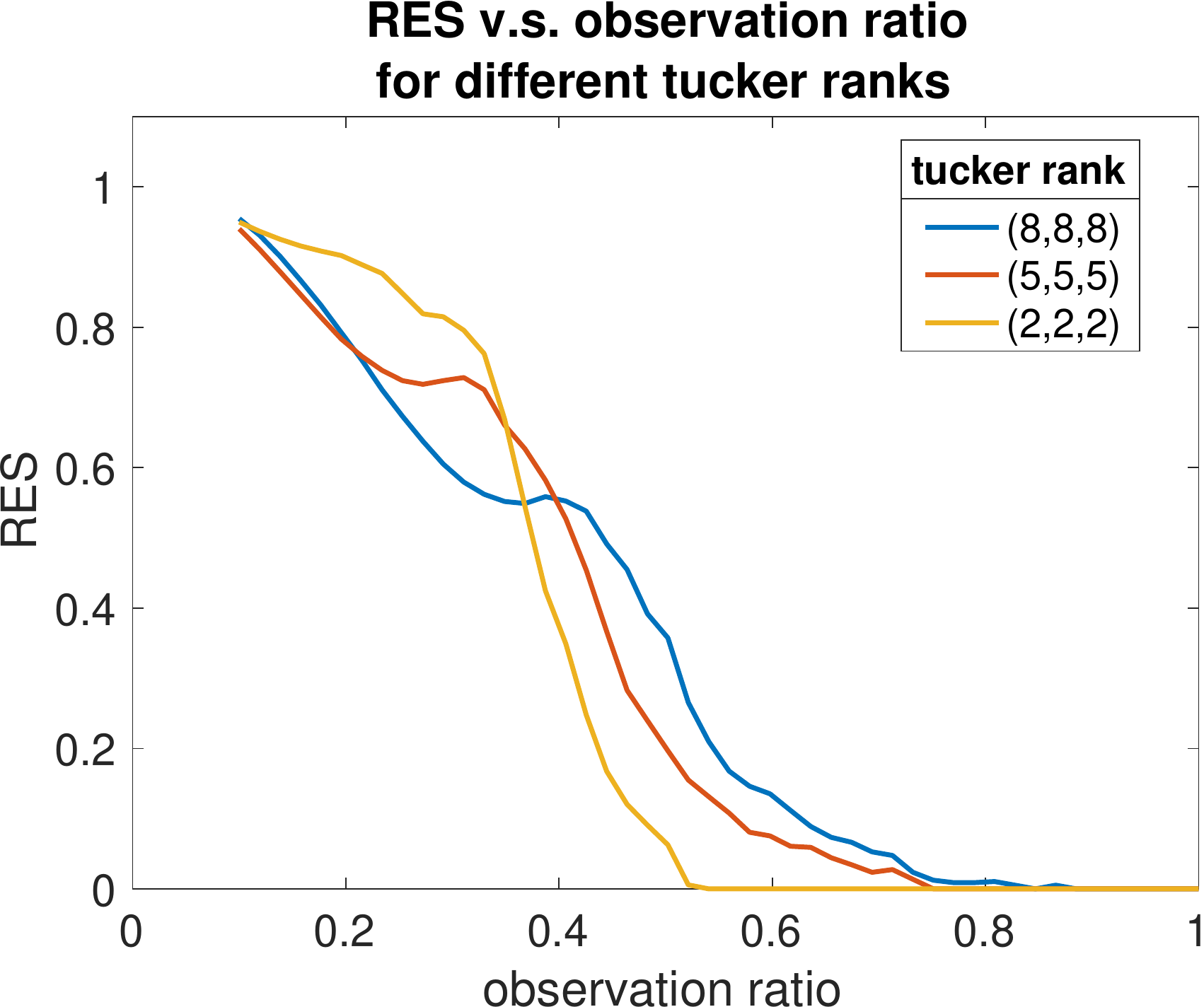}}
\hfil
\subfloat[]{\includegraphics[width=0.45\textwidth]{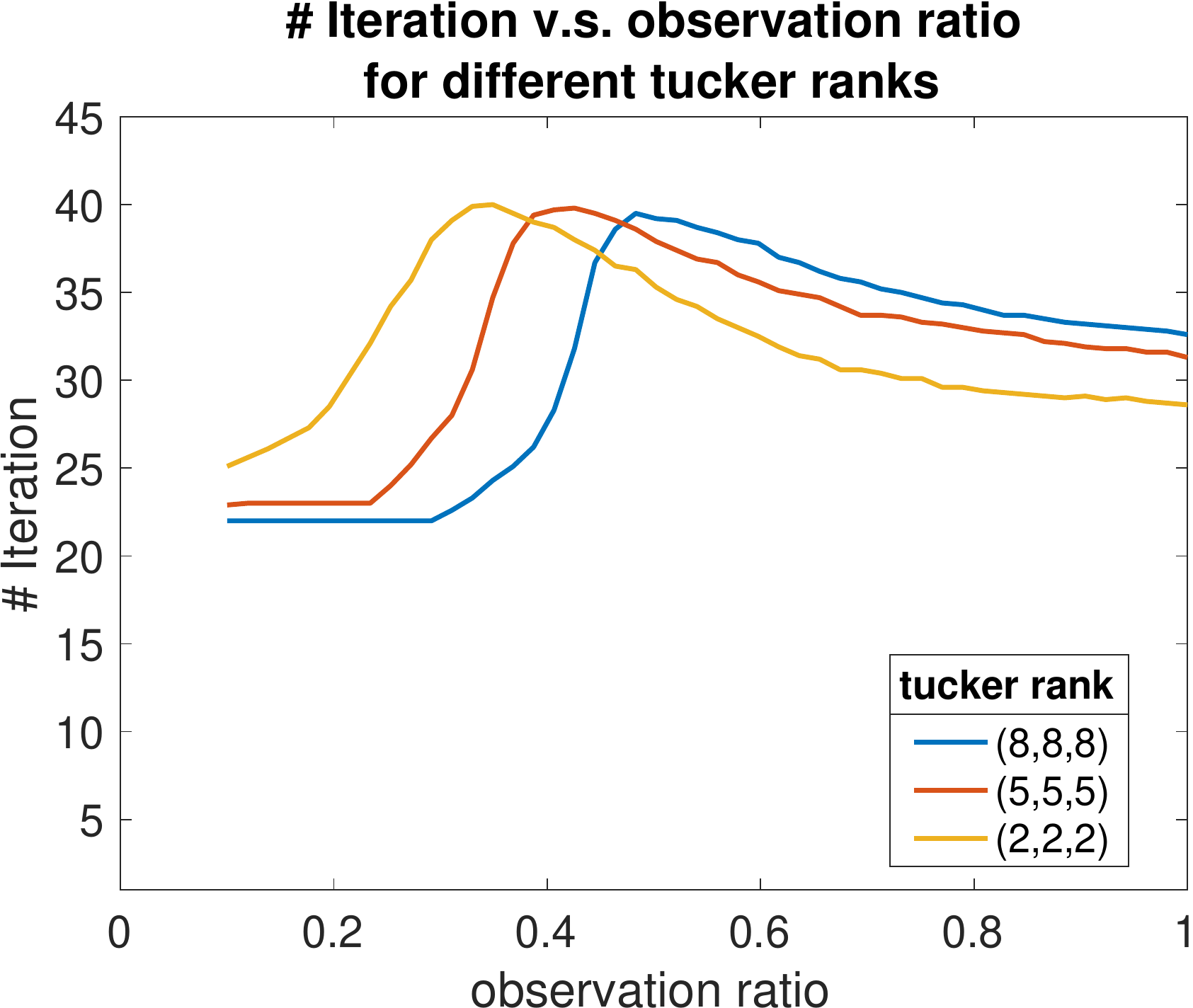}}
\hfil
\subfloat[]{\includegraphics[width=0.45\textwidth]{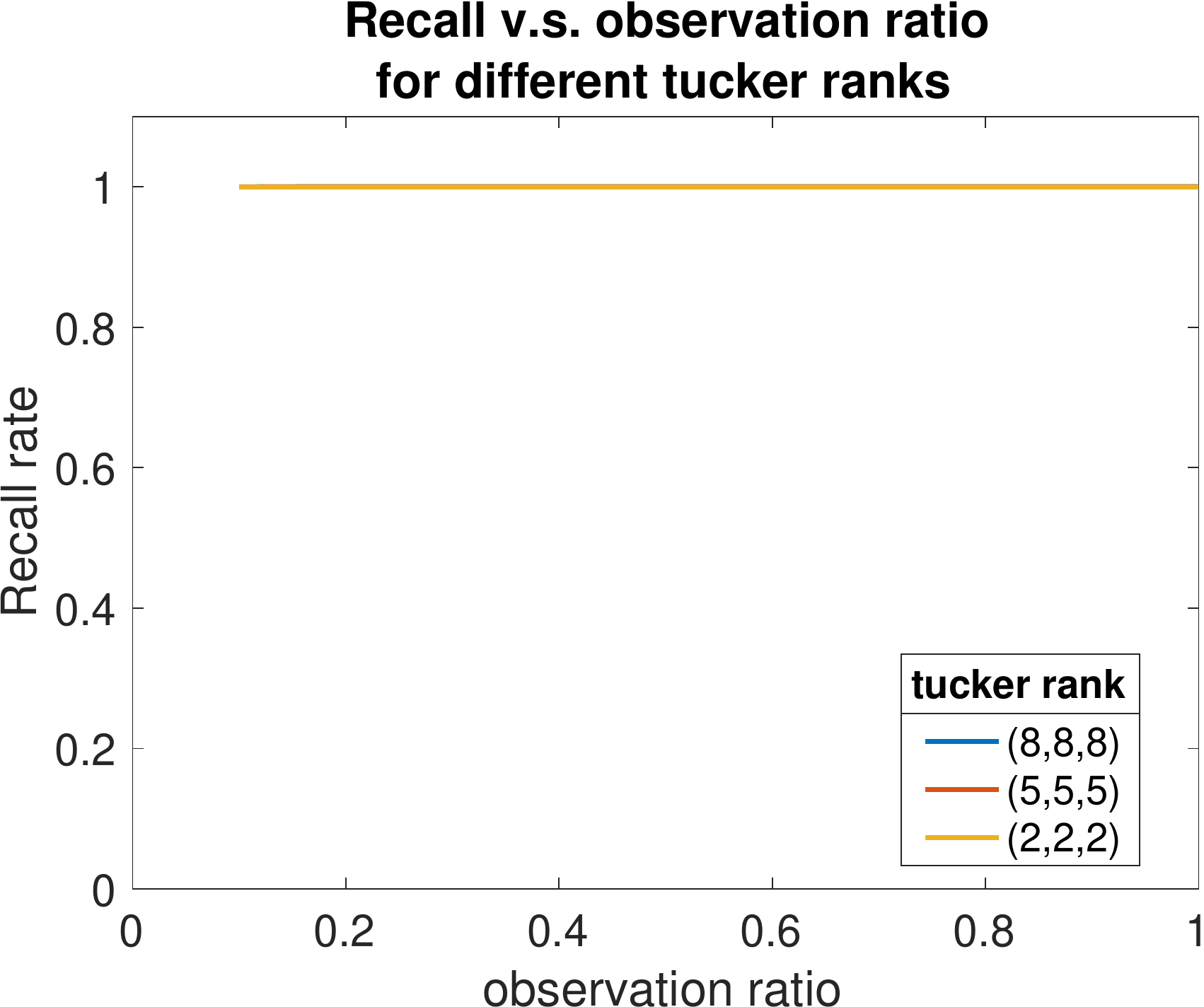}} 
\hfil
\subfloat[]{\includegraphics[width=0.45\textwidth]{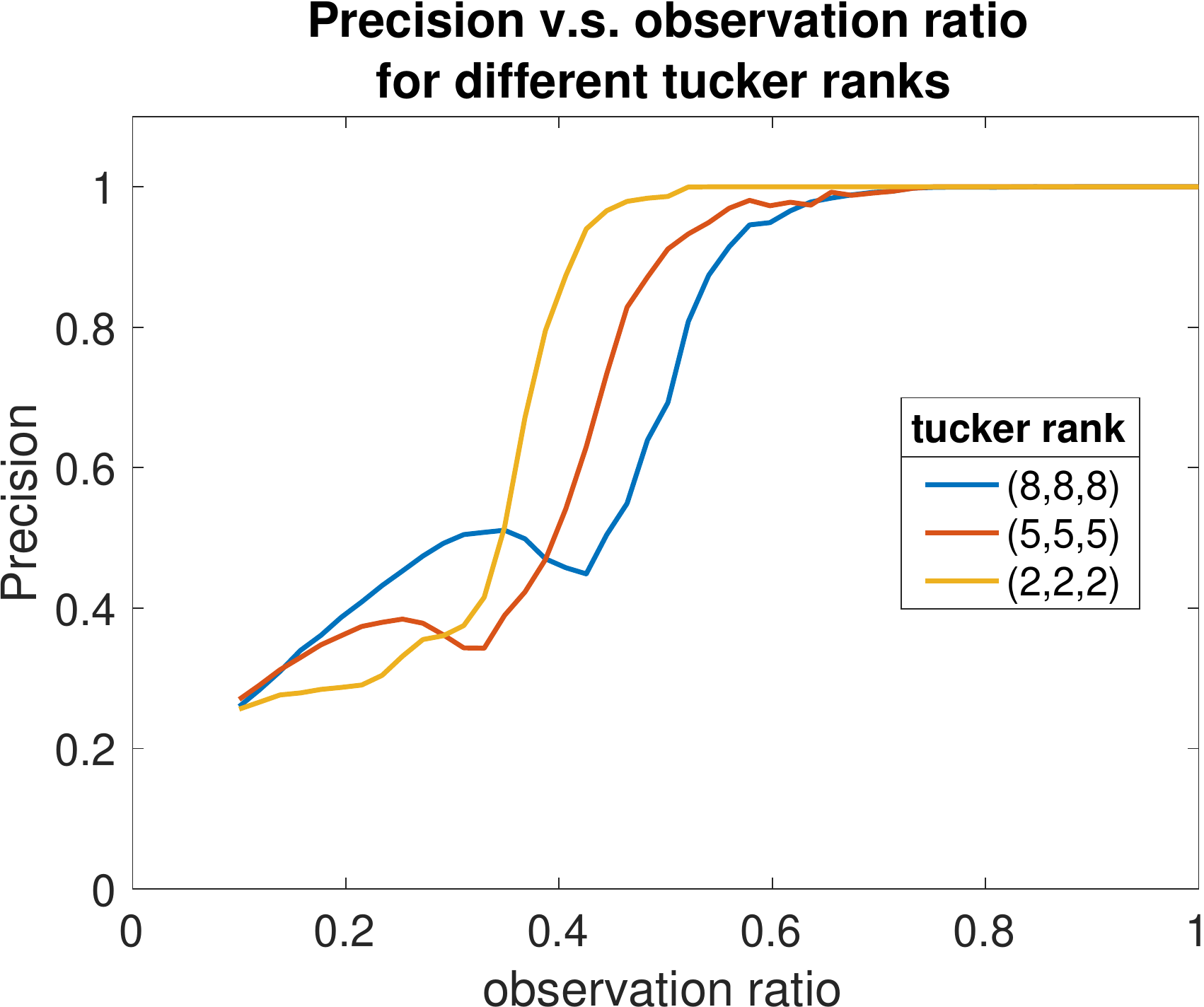}}
\caption{Results of Algorithm~\ref{alg:TC} as a function of the observation ratio considering $\mathcal{X}_0$  with varying tucker rank. The low-rank tensor $\mathcal{X}_0$ is generated with a tucker rank of (2,2,2), (5,5,5), and (8,8,8) respectively, with a fixed size of $\mathbb{R}^{70 \times 70 \times 70}$, and a gross corruption ratio $\gamma = 0.1$. The plotted result is an average over 10 trials.}
\label{fig:TC_rank}
\end{figure}

Next we fix the low-rank tensor $\mathcal{X}_0$ at size $  \mathbb{R}^{70 \times 70 \times 70}$ and gross corruption ratio $\gamma = 0.1$. For $\mathcal{X}_0$ of different tucker ranks (2,2,2),(5,5,5), and (8,8,8), we vary the observation ratio from 0.1 to 1 and run Algorithm \ref{alg:TC}.  The result is shown in Figure \ref{fig:TC_rank}, which is an average across 10 trials. Again, we observe a phase-transaction behavior, that when observation ratio is above a critical threshold, the decomposition is exact, with precision and recall at one, and relative error at zero. For tensor ranks (5,5,5) and (8,8,8), this threshold is about 0.8. For tensor rank (2,2,2), it is lower, about 0.5. This indicates that when the underlying tensor rank is lower, we can exactly conduct the decomposition with even with a lower observation ratio.

\subsubsection{Comparison with $l_{2,1}$ regularized tensor decomposition} Next, we compare the performance of Algorithm~\ref{alg:TC} ($l_{2,1}$ norm regularized tensor completion), with $l_1$ norm regularized tensor completion. We fix the low-rank tensor $\mathcal{X}_0$ at size $\mathbb{R}^{70 \times 70 \times 70}$ with a tucker rank of $(5,5,5)$, and fix gross corruption ratio  $\gamma = 0.1$. We vary the observation ratio $\rho$ from 0.1 to 1 and run Algorithm \ref{alg:TC} and $l_1$ norm regularized tensor completion. We run 10 times and average the results.

\begin{figure}
\centering
\subfloat[]{\includegraphics[width=0.45\textwidth]{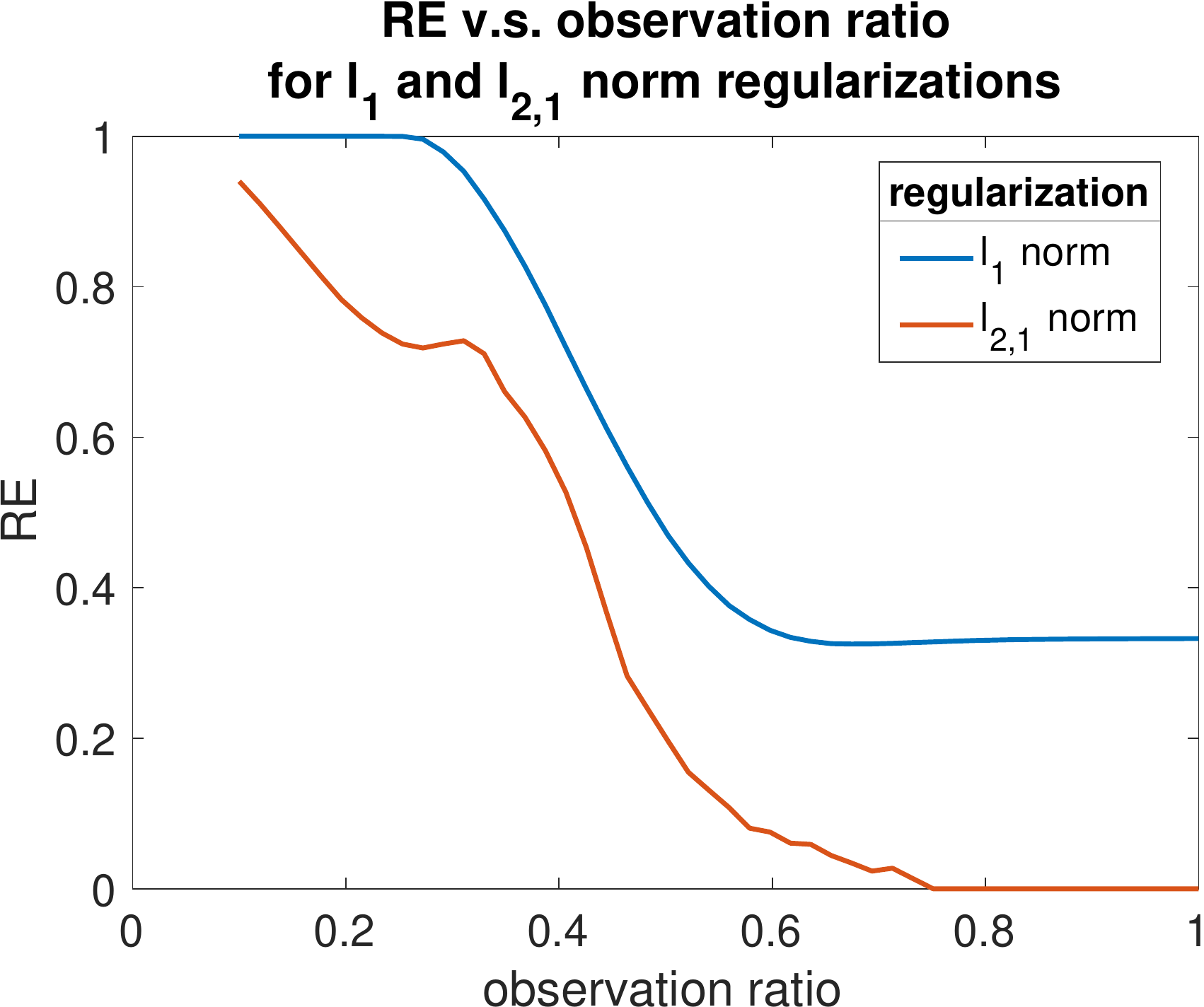}}
\hfil
\subfloat[]{\includegraphics[width=0.45\textwidth]{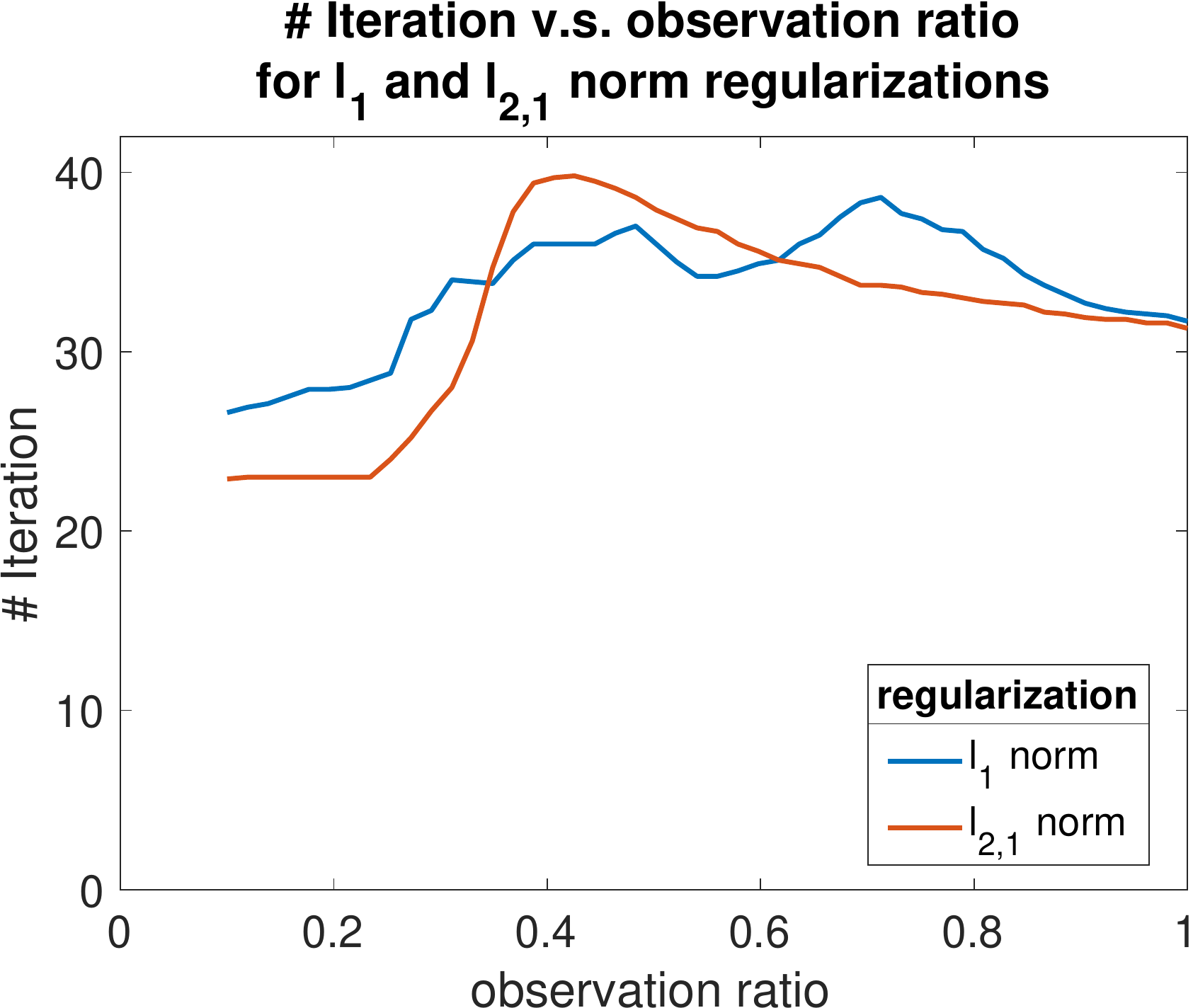}}
\hfil
\subfloat[]{\includegraphics[width=0.45\textwidth]{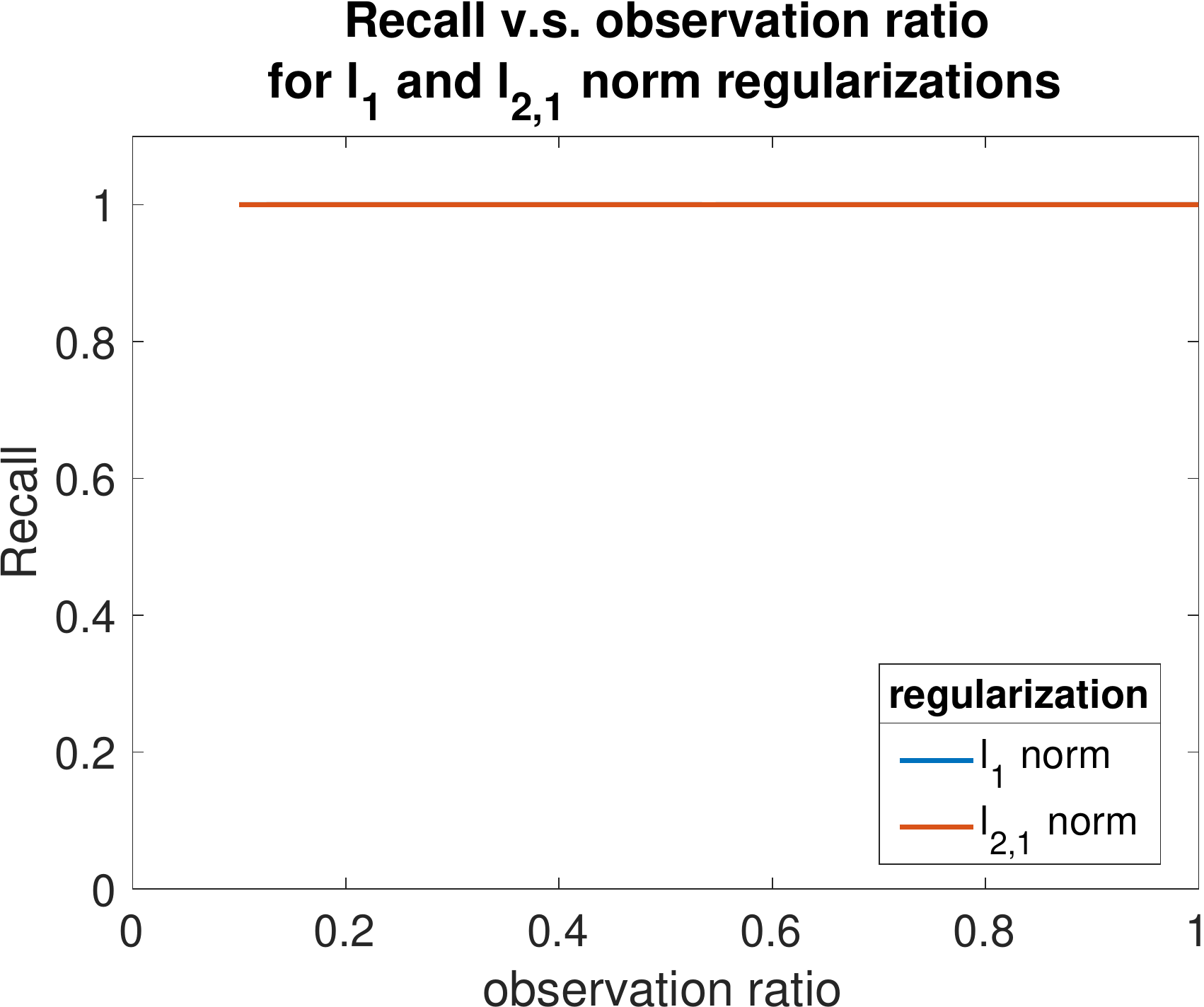}} 
\hfil
\subfloat[]{\includegraphics[width=0.45\textwidth]{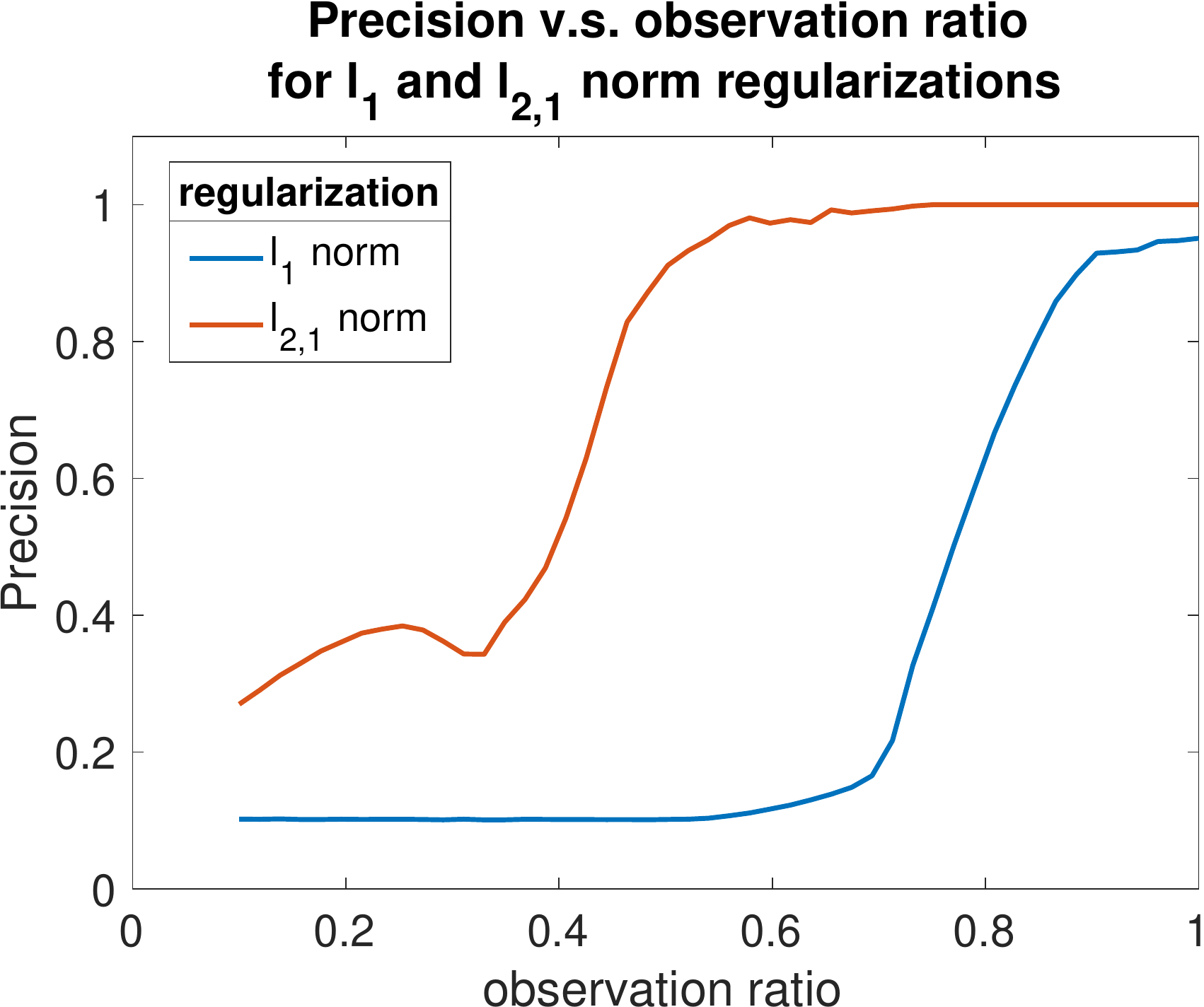}}
\caption{Comparison of Algorithm~\ref{alg:TC} ($l_{2,1}$ norm regularized tensor completion), with $l_1$ norm regularized tensor completion, as observation ratio varies.  The low-rank tensor $\mathcal{X}_0$ size is fixed at $\mathbb{R}^{70 \times 70 \times 70}$ with a tucker rank of $(5,5,5)$, and the gross corruption ratio $\gamma$ is set at 0.1. The result is an average over 10 trials.}
\label{fig:TC_compare}
\end{figure}

Figure \ref{fig:TC_compare} shows the result. We can see that the performance under $l_1$ norm regularization is constantly worse, with a higher relative residual error and a lower precision. With a corruption ratio of 0.1, even when data is fully observed,  $l_1$ norm regularized tensor completion can only achieve a relative error of around 0.3, and a precision of around 0.9. Moreover, $l_1$ norm regularization has a smaller range of observation ratio, outside of which the performance drops sharply, namely $\rho > 0.9$, compared with approximately $\rho > 0.7$ for $l_{2,1}$ norm regularization.

\subsubsection{Phase transition behavior} 
\begin{figure} 
\centering
\includegraphics[width=0.6\linewidth]{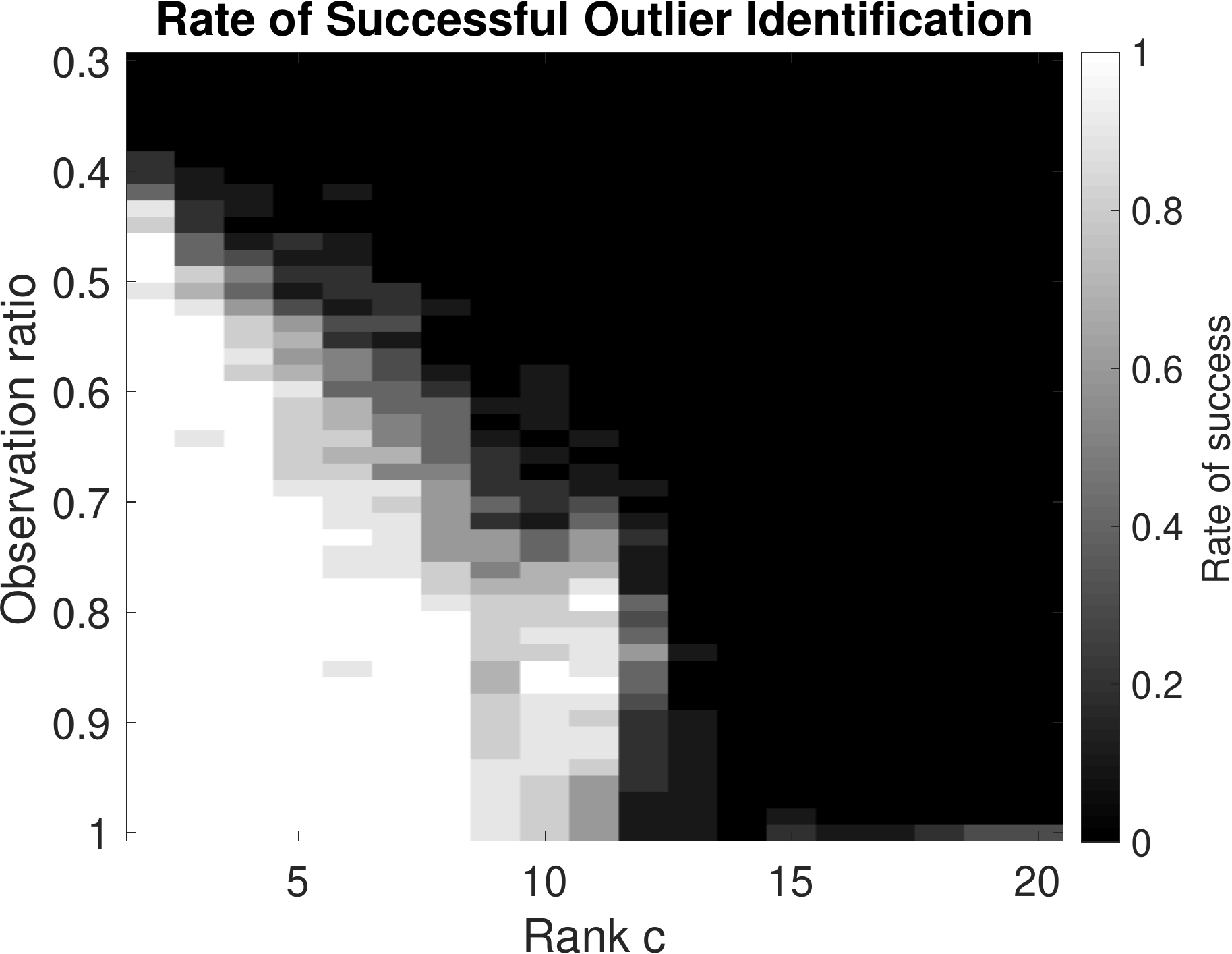}
\caption{Rate of successful outlier identification across 10 trials. The color denotes the rate of success. For each trial we create a tensor with size $\mathbb{R}^{70 \times 70 \times 70}$ and tucker rank (c,c,c) (x-axis), fiber-wise corrupt it at ratio $\gamma = 0.1$. We vary the observation ratio $rho$ from 0.3 to 1 (y-axis).}
\label{fig:block}
\end{figure}

Now, we further study the phase transition property of Algorithm \ref{alg:TC} in terms of the observation ratio and the tensor rank. We fix the gross corruption ratio at $\gamma = 0.1$ and the tensor size at $  \mathbb{R}^{70 \times 70 \times 70}$. Then we vary the observation ratio from 0.3 to 1, and the tucker rank of $\mathcal{X}_0$ from (1,1,1) to (20,20,20). For each combination we conduct 10 trials. Figure \ref{fig:block} shows the success rate out of 10 trials for varying tensor rank and observation ratio. We regard a trial \textit{successful} if both the precision and recall of the outlier location identification are greater than 0.99. The result shows that the possibility of success rises as observation ratio increases and tucker rank of $\mathcal{X}_0$ decreases. For observation ratios greater than 0.7 and tucker rank smaller than (5,5,5), the outlier identification is always successful.

\section{Case study: Nashville, TN traffic dataset} \label{Sec:case}
In this section, we apply our proposed method to a dataset of real traffic data  and use it to detect traffic events. We use the traffic speed data of downtown Nashville from Jan 1 to Apr 29, 2018 obtained from a large scale traffic aggregator. Given that this is a real empirical dataset, we do not have access to the true low rank traffic conditions and the true outliers. As a consequence, it is not possible to evaluate the precision and recall of the outlier detection algorithm as was done in the numerical examples in the previous section. Nevertheless, our algorithm can mark the events that are confirmed to be severe car crashes, construction lane closures, or large events that caused significant disruption on traffic of downtown Nashville.

\begin{figure} 
\centering
\includegraphics[width=0.6\linewidth]{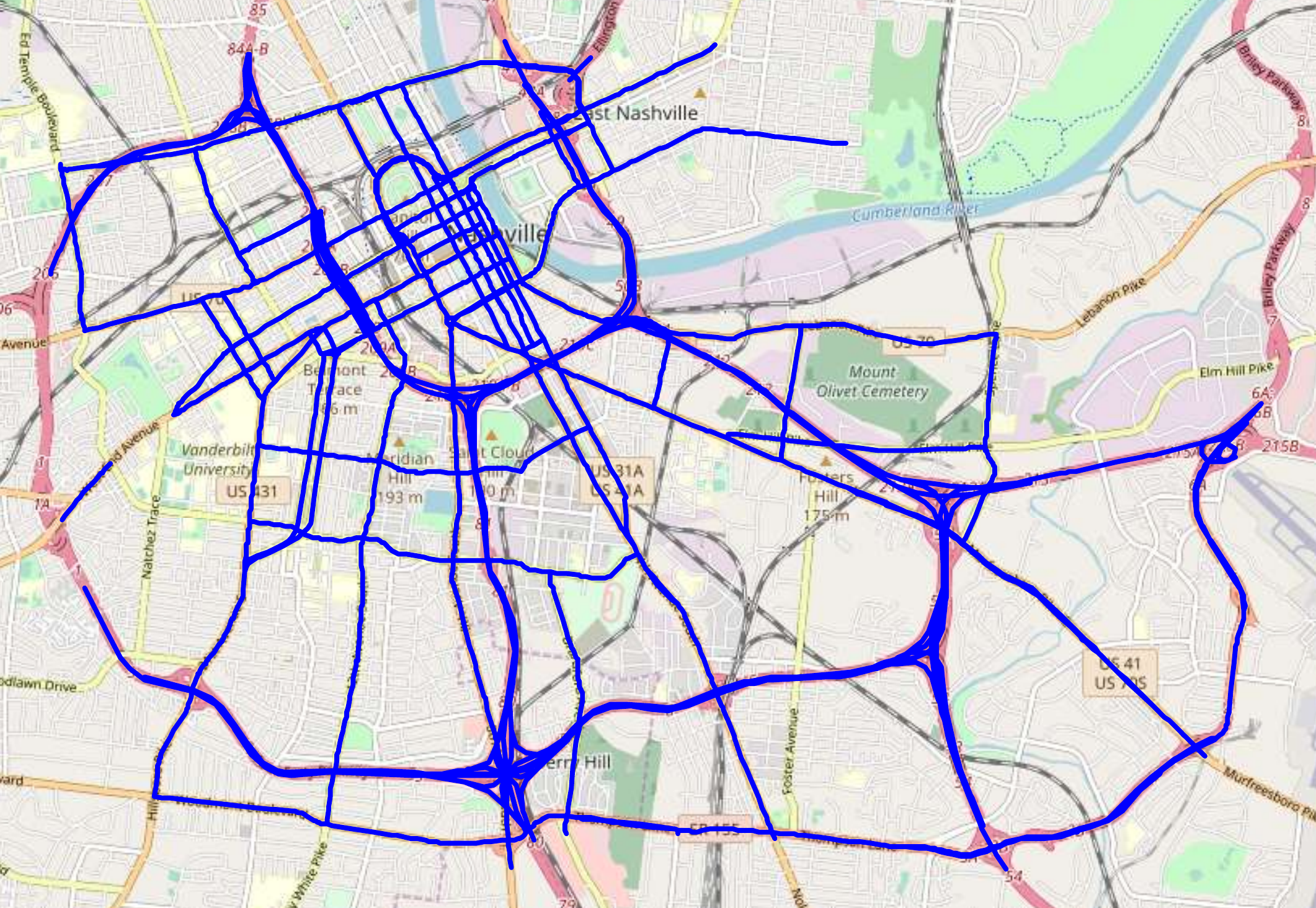}
\caption{Map of downtown Nashville. The studied road segments are marked in blue and consist of the major freeways and surface streets.}
\label{fig:downtown_map}
\end{figure}

We select a subset of road segments within downtown Nashville area that regularly have traffic data available. The base traffic dataset consists of the one-hour average speed of traffic on each road segment in the network. 
The dataset has an observation ratio of 0.807, and records 556 road segments for 17 weeks, every week containing $24 \times 7 = 168$ hours, for a total of 2856 hours. We can thus construct a data tensor of size $556 \times 168 \times 17$. The map of the final-selected road segments are shown in Figure \ref{fig:downtown_map}, containing major interstate highways I-40, I-24 and I-440, among other major surface streets. 

\begin{figure} 
\centering
\includegraphics[width=0.7\linewidth]{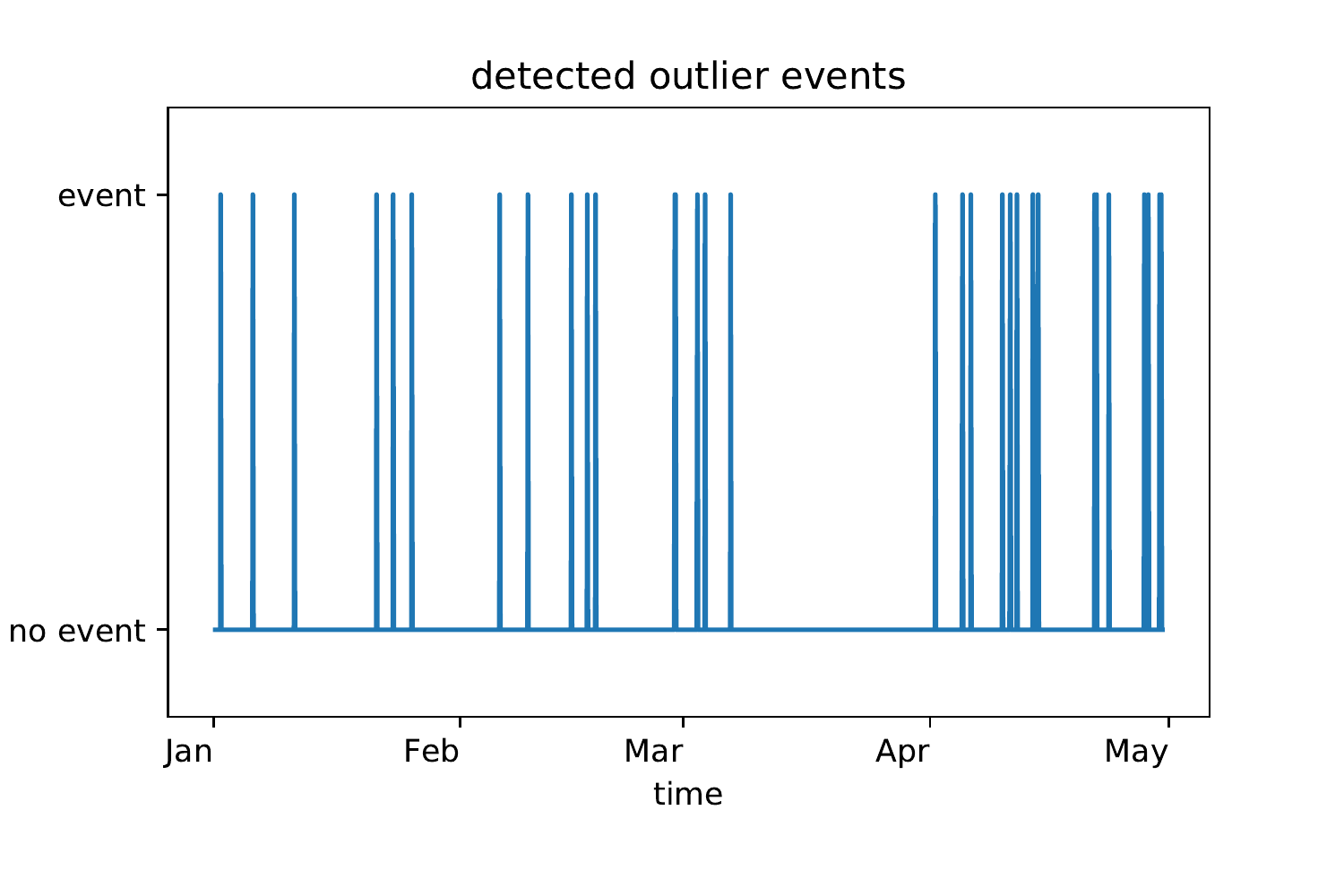}
\caption{Stem plot of detected outlier events. Each column indicates the time when an detected outlier event takes place.  }
\label{fig:dates_ano}
\end{figure}

Since the data is not fully observed, we adopt the robust tensor completion algorithm (Algorithm~\ref{alg:TC}). We set $\lambda = 1.47$, leading to a corruption ratio of 1.18\%. This means over the 17 weeks (2856 hours), 36 hours are marked as abnormal. Algorithm~\ref{alg:TC} takes 19 seconds to run on this dataset.
Figure \ref{fig:dates_ano} plots the timeline when these outlier events take place.

Next, we investigate the outlier events identified by Algorithm \ref{alg:TC}. Out of the 36 hours detected as abnormal, 31 can be easily matched to recorded incidents. This includes construction lane closures, car crashes, and large events like the annual St. Jude Rock ‘N’ Roll Marathon~\cite{marathon2018}. This process is done by manually comparing the events identified by Algorithm~\ref{alg:TC} with the accident records of the Nashville fire department~\cite{firedpm}, and lane closure records of the\textit{Tennessee Department of Transportation} (TDOT)~\cite{TDOT2018}, which is the state transportation authority. Most incidents clear out after one hour, while some last for two hours or more.  For the rest 5 hours detected as outliers, the average speed of the roads appear faster than normal, and we are not able to identify obvious causes for the abnormally light traffic conditions. 

\begin{figure} 
\centering
\includegraphics[width=\textwidth]{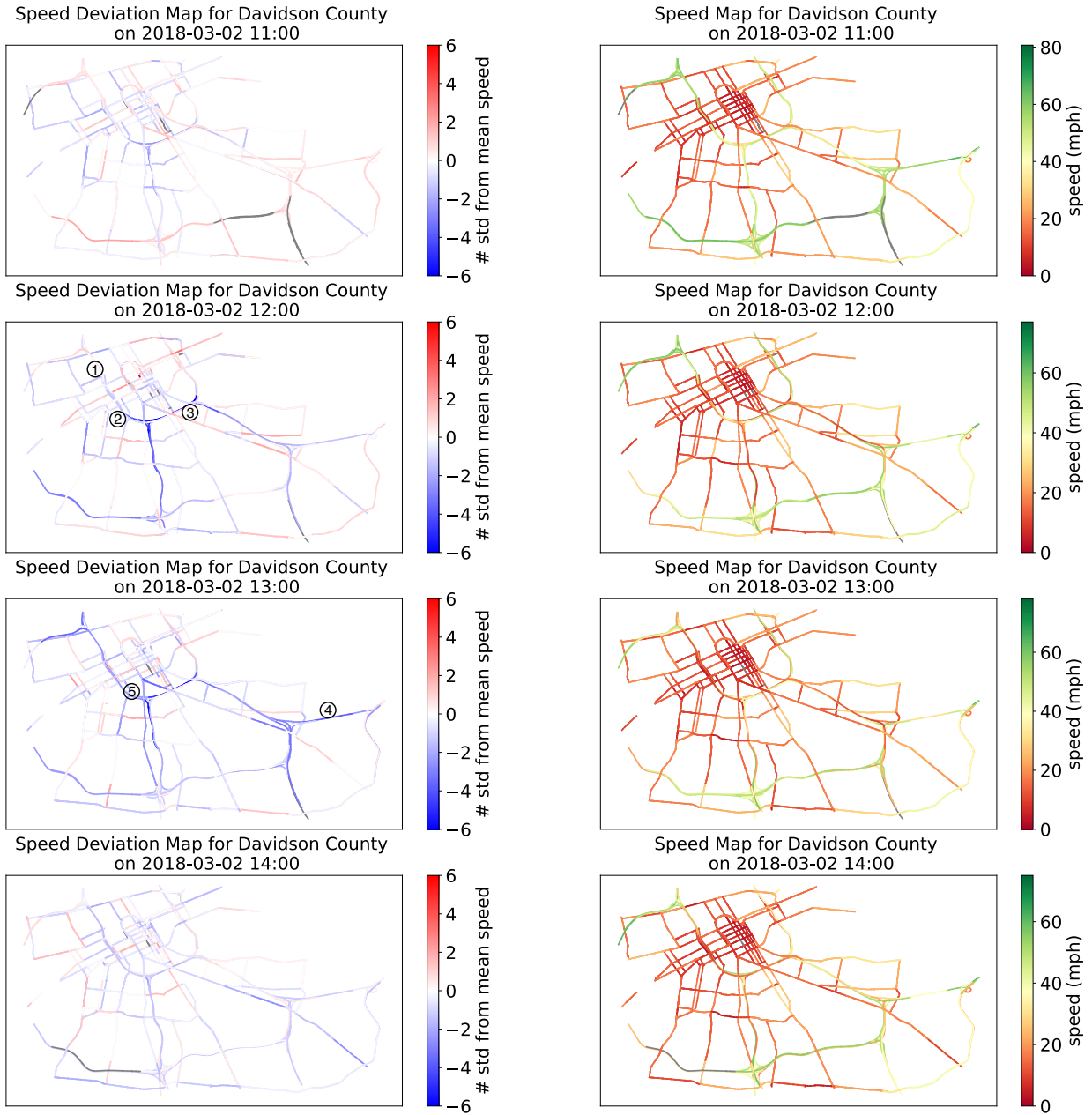}
\caption{Detected car crashes. The second and third row, i.e., 12:00 and 13:00 are marked as outliers. At 12:00, several severe car crashes happened on Interstate 40 (area 1 in the second row), Charlotte Avenue (area 2 in the second row), and Carroll street (area 3 in the second row). At about 13:00, two other car crashes occur at different segments of Interstate 40 (areas 4 and 5 in the third row).  }
\label{fig:Crush}
\end{figure}

\begin{figure} 
\centering
\includegraphics[width=\textwidth]{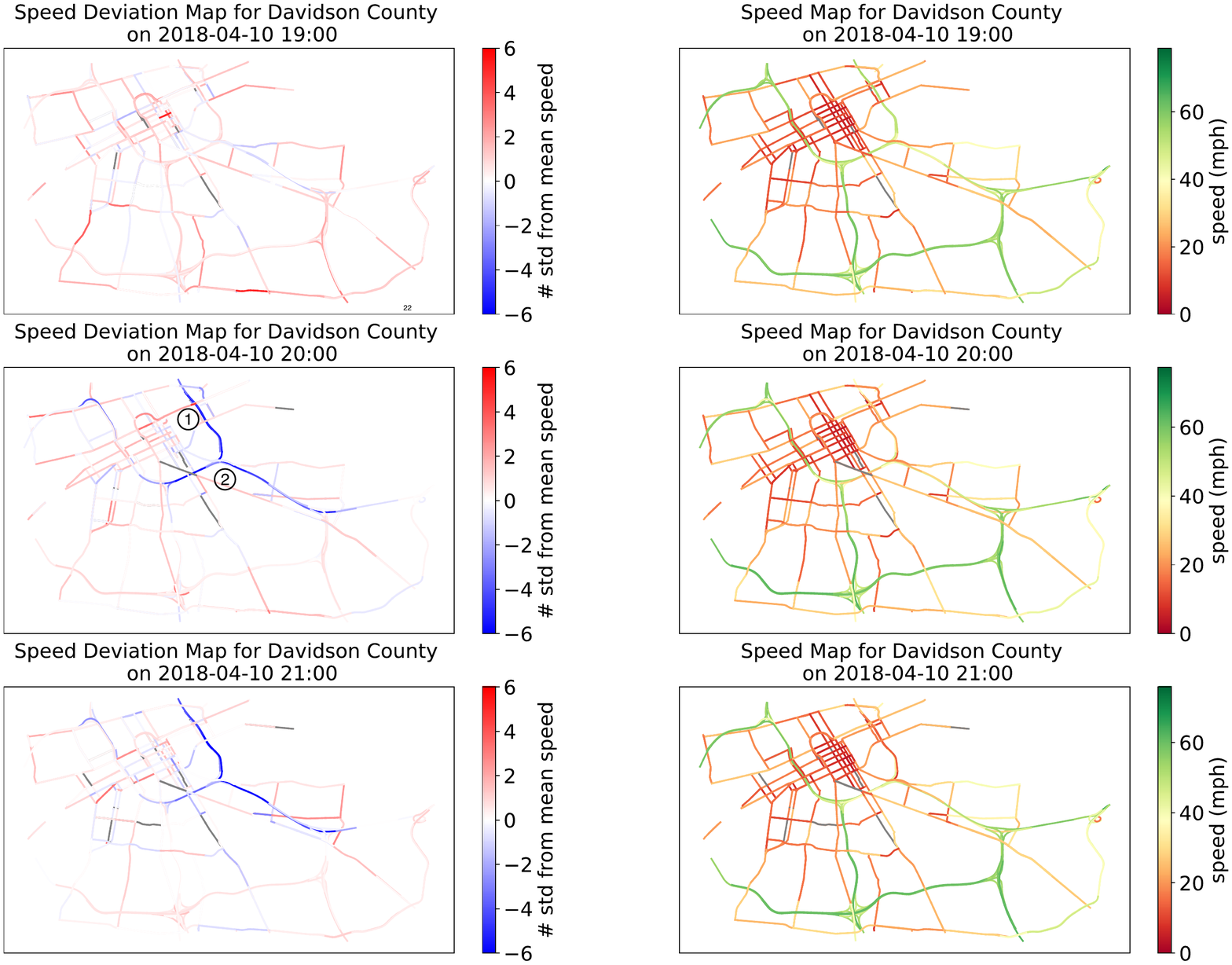}
\caption{Detected road closure. The second and third row, i.e. 20:00 and 21:00 are marked as outliers. It is the time when segments of Interstate 24 (the north-south route marked as 1), Interstate 40 (the east-west route marked as 2), as well as streets connecting to Interstate 24, were closed for bridge rehabilitation, resurfacing, and maintenance.}
\label{fig:Construct}
\end{figure}

Figures~\ref{fig:Crush} and~\ref{fig:Construct} visualize some outlier events as examples. The right columns of the heat maps show the average speed of the road in one hour, and the left columns show how many standard deviations the road segment is from from the average speed of that hour. We calculate the average speed by looking at the low rank matrix $\hat{\mathcal{X}}$ of Algorithm \ref{alg:TC}, which is expected to be the normal traffic pattern, and calculate the mean speed of 17 weeks for every hour of the week.

 Figure \ref{fig:Crush} shows an event that corresponds to a series of car crashes. On 12:00 noon March 2, 2019 several severe car crashes happened on the major freeway and arterial going around downtown Nashville, namely Interstate 40 (labeled 1 in the second row of Figure~\ref{fig:Crush}), Charlotte Avenue (labeled 2 in the second row), and Carroll street (labeled 3 in the second row). At about 13:00, two other car crashes happened in different segments of Interstate 40 (labeled 4 and 5 in the third row of~Figure~\ref{fig:Crush})~\cite{firedpm}. The sequence of severe crashes created unusual congestion for that time of day and took two hours to clear out. The Algorithm marks the the hours 12:00 and 13:00 as outliers. 
 
 Figure~\ref{fig:Construct} shows a detected construction event. From 20:00 Tuesday evening through 5:00 the following day, there were road closures for bridge rehabilitation, resurfacing and maintenance on the major freeways around downtown Nashville, namely Interstate 24 (the north-south route marked as 1), Interstate 40 (the east-west route marked as 2), as well as streets connecting to Interstate 24~\cite{TDOTApril}. The first two hours of the lane closure, i.e., 20:00 and 21:00, observed the most congestion and were detected as outliers. The late night hours of Tuesday evening and the early Wednesday morning hours did not experience significant congestion and were consequently not detected as outliers.


\section{Conclusions}
In this work, we introduced a tensor completion problem to detect extreme traffic conditions that exploits the spatial and temporal structure of traffic patterns in cities. An algorithm is proposed to perform the detection even in the presence of missing data. The method was applied to numerical examples that demonstrate exact recovery of the underlying low rank tensor is possible in a range of settings with corrupted and missing entries, with lower quality results achieved as the fraction of missing entries increases. A case study on traffic conditions in Nashville, TN, demonstrates the practical performance of the method.

One limitation of the proposed approach is that the method exploits linear relationships between the traffic patterns, and is not designed to capture nonlinear spatial and temporal relationships. In our future work we are interested in exploring possible neural network extensions might generalize the outlier detection tools for more complex relationships.

\section{Acknowledgement}
This material is based upon work supported by the National Science Foundation under Grant No.
CMMI-1727785.

\bibliographystyle{unsrt}
\bibliography{ref}
\end{document}